\documentclass[lettersize,journal]{IEEEtran}
\usepackage{amsmath,amsfonts}
\usepackage{algorithmic}
\usepackage{algorithm}
\usepackage{array}
\usepackage{textcomp}
\usepackage{stfloats}
\usepackage{url}
\usepackage{verbatim}
\usepackage{cite}
\usepackage{amsthm}
\usepackage{amsfonts}
\usepackage{amsmath} 
\usepackage{physics} 

\usepackage{graphicx}
\usepackage{epstopdf}

\usepackage{amssymb}

\usepackage{color} 
\usepackage{bm}
\usepackage{epstopdf} 
\usepackage{booktabs}  
\usepackage[dvipsnames]{xcolor} 
\usepackage{cite} 
\usepackage{balance}
\usepackage{setspace}  
\usepackage{multirow} 
\usepackage{makecell} 
\usepackage{enumerate} 
\usepackage{subfigure} 
\usepackage{hyperref}
\usepackage {wrapfig}
\usepackage{mathtools}

\DeclareMathAlphabet{\mathsfit}{\encodingdefault}{\sfdefault}{m}{sl}
\SetMathAlphabet{\mathsfit}{bold}{\encodingdefault}{\sfdefault}{bx}{n}
\newcommand{\tens}[1]{\bm{\mathsfit{#1}}}
\def\tA{{\tens{A}}}
\def\tB{{\tens{B}}}

\def\tH{{\tens{H}}}

\def\tU{{\tens{U}}}

\def\tW{{\tens{W}}}
\def\tX{{\tens{X}}}
\def\tY{{\tens{Y}}}

\newtheorem{ppn}{Proposition}

\newtheorem*{pf}{Proof}  
\def\widebreve{\mathpalette\wide@breve}
\def\wide@breve#1#2{\sbox\z@{$#1#2$}%
	\mathop{\vbox{\m@th\ialign{##\crcr
				\kern0.08em\brevefill#1{0.8\wd\z@}\crcr\noalign{\nointerlineskip}%
				$\hss#1#2\hss$\crcr}}}\limits}
\def\brevefill#1#2{$\m@th\sbox\tw@{$#1($}%
	\hss\resizebox{#2}{\wd\tw@}{\rotatebox[origin=c]{90}{\upshape(}}\hss$}

\newcommand{\ppnref}[1]{\textbf{Proposition \ref{#1}}}

\newcommand{\figref}[1]{Fig. \ref{#1}}

\allowdisplaybreaks[4] 

\makeatletter
\renewcommand{\maketag@@@}[1]{\hbox{\m@th\normalsize\normalfont#1}}%
\makeatother

\begin{document}
\title{Towards Unified AI Models for MU-MIMO Communications: A Tensor Equivariance Framework}

	\author{Yafei Wang, \textit{Graduate Student Member}, \textit{IEEE}, Hongwei Hou, \textit{Graduate Student Member}, \\Xinping Yi, \textit{Member}, \textit{IEEE}, Wenjin Wang, \textit{Member}, \textit{IEEE}, Shi Jin, \textit{Fellow}, \textit{IEEE}
	\thanks{The codes are open-sourced on \url{https://github.com/ZJSXYZ/TENN}.}
	\thanks{Manuscript received xxx; revised xxx. The associate editor coordinating the review of this article and approving it for publication was xxx. \textit{(Yafei Wang and Hongwei Hou contributed equally to this work.)} \textit{(Corresponding author: Wenjin Wang.)}}
	\thanks{Yafei Wang, Hongwei Hou, and Wenjin Wang are with the National Mobile Communications Research Laboratory, Southeast University, Nanjing 210096, China, and also with Purple Mountain Laboratories, Nanjing 211100, China (e-mail: wangyf@seu.edu.cn; hongweihou@seu.edu.cn; wangwj@seu.edu.cn).}
	\thanks{Xinping Yi and Shi Jin are with the National Mobile Communications Research Laboratory, Southeast University, Nanjing 210096, China (e-mail: xyi@seu.edu.cn; jinshi@seu.edu.cn).}}
	
	
	\maketitle
	
	\begin{abstract}
		In this paper, we propose a unified framework based on equivariance for the design of artificial intelligence (AI)-assisted technologies in multi-user multiple-input-multiple-output (MU-MIMO) systems. We first provide definitions of multidimensional equivariance, high-order equivariance, and multidimensional invariance (referred to collectively as tensor equivariance). On this basis, by investigating the design of precoding and user scheduling, which are key techniques in MU-MIMO systems, we delve deeper into revealing tensor equivariance of the mappings from channel information to optimal precoding tensors, precoding auxiliary tensors, and scheduling indicators, respectively. To model mappings with tensor equivariance, we propose a series of plug-and-play tensor equivariant neural network (TENN) modules, where the computation involving intricate parameter sharing patterns is transformed into concise tensor operations. Building upon TENN modules, we propose the unified tensor equivariance framework that can be applicable to various communication tasks, based on which we easily accomplish the design of corresponding AI-assisted precoding and user scheduling schemes.  
		Simulation results show that the proposed methods achieve near-optimal performance with significantly lower complexity and strong generalization across multiple dimensions. For instance, the NN trained for precoding with 8 users provides satisfactory performance in a 10-user scenario.
		This validates the superiority of TENN modules and the unified framework.
	\end{abstract}
	
	
	\bstctlcite{IEEEexample:BSTcontrol} 
		\section{Introduction}
	\IEEEPARstart{T}{he} multiple-input-multiple-output (MIMO) technology \cite{6375940}, which serves users using multiple antennas, has become a key technology in wireless communication systems due to its enormous potential for increasing system capacity \cite{wang2023road}. Leveraging MIMO technology, the base stations (BSs) in multi-user MIMO (MU-MIMO) systems possess enhanced transmission performance to simultaneously serve multiple users. In such systems, resource allocation schemes such as user scheduling and precoding techniques play an important role in improving throughput. On the one hand, user scheduling aims to select users from the pool of all users for simultaneous transmission in certain resource elements, with the goal of improving transmission quality. On the other hand, precoding further enhances the potential capacity gains by suppressing user interference. Since the evolution of MU-MIMO technology, numerous excellent scheduling and precoding algorithms have been proposed, such as greedy-based scheduling schemes \cite{dimic2005downlink} and weighted minimum mean square error (WMMSE) precoding \cite{4712693, shi2011iteratively,zhao2023rethinking}.
	
	While conventional transmission schemes in MU-MIMO systems can achieve outstanding performance \cite{4712693, dimic2005downlink}, even approaching the performance limits \cite{1413598, 1056659}, they usually require iterative computations and high computational complexity. Such issues become increasingly severe as the growing scale of wireless communication systems \cite{8323218}, posing obstacles to their application in practical systems. 
	In contrast, artificial intelligence (AI) models possess the potential to accelerate iteration and approximate high-dimensional mappings with lower computational complexity \cite{hornik1989multilayer,yun2019transformers}, leading to extensive research into AI-assisted transmission schemes \cite{letaief2019roadmap}.
	
	
	Most AI-assisted transmission schemes treat inputs as structured data, such as image data or vector data, and process them using corresponding neural networks (NNs) \cite{kim2020deep, huang2019fast, 9322310, 8935405, 9516008, shi2023robust}. Specifically, building on the optimal closed-form solution derived from WMMSE precoding \cite{4712693, 6832894}, the authors in \cite{kim2020deep} utilize fully connected (FC) layers to directly compute key features in optimal solution forms, while similar schemes are proposed with convolutional NN (CNN) in \cite{huang2019fast} and \cite{9322310}, as channel information can be regarded as image data. Based on CNN, AI-assisted schemes utilizing optimal solution structures have been extended to multiple precoding optimization criteria \cite{8935405}. 
	With the imperfect channel state information (CSI), \cite{9516008} and \cite{shi2023robust} investigated the robust WMMSE precoding algorithms with CNNs. 
	Unlike FC and CNN networks, deep unfolding networks integrate learnable parameters into iterative algorithms to expedite algorithmic convergence \cite{9667094, 9246287, wang2024robust}. For instance, \cite{9667094} introduced a matrix-inverse-free deep unfolding network for WMMSE precoding. A similar approach is explored in \cite{9246287}, demonstrating better performance compared to CNNs. Such approach is further extended to WMMSE precoding design under imperfect CSI conditions \cite{wang2024robust}. 
	Apart from precoding, there are also some AI-assisted methods for other resource allocation schemes \cite{8922744, li2021user, xie2024learning}.
	In \cite{8922744}, FC networks are utilized to extract optimal power allocation schemes from CSI for maximizing sum-rate. Besides, a user scheduling strategy aided by FC networks is proposed in \cite{li2021user}, which assigns the most suitable single user for each resource block. Based on edge cloud computing and deep reinforcement learning, the user scheduling strategy in \cite{xie2024learning} are developed for millimeter-wave vehicular networks. Additionally, there are studies providing novel ideas to understand and assist in the design of AI-assisted methods in wireless communications \cite{sun2018learning, 9674231}. The work in \cite{sun2018learning} elucidates the shortcomings of AI in solving non-convex problems and presents a framework to address this issue. The authors of \cite{9674231} investigated the asymptotic spectral representation of linear convolutional layers, offering guidance on the excellent performance of CNNs.
	
	The aforementioned studies typically do not focus on permutation equivariance (abbreviated as equivariance) \cite{zaheer2017deep}, which entails that permutation of input elements in a model also results in the corresponding permutation of output elements. Such property is inherent to MU-MIMO systems and endows AI-assisted transmission technologies with the potential advantages like parameter sharing \cite{ravanbakhsh2017equivariance}. Benefiting from its modeling of graph topology, graph NN (GNN) possess the capability to exploit this property, thus being employed in the design of transmission schemes and demonstrating outstanding performance \cite{shen2022graph}.
	In \cite{yi2015topological}, the significance of topological information for transmission within an interference management framework is investigated. The GNNs used for wireless resource management is proposed in \cite{shen2020graph}, which develops equivariance and thereby achieves generalization across varying user count. In addition, the authors in \cite{kim2022bipartite} model the link network between BS antennas and terminals as a bipartite graph, thereby achieving generalization across varying numbers of users and BS antennas. Similarly, GNNs with different iteration mechanisms are proposed for precoding design in \cite{guo2021learning} and \cite{zhao2022learning}. By crafting refined strategies for updating node features, a GNN satisfying equivariance across multiple node types is devised for hybrid precoding in \cite{liu2023multidimensional}. This work also reveals various potential scenarios and tasks for the GNN method. The proposed methodology demonstrates exceptional performance and scalability, paving the way for GNN-assisted transmission design. Furthermore, aiming to maximize the number of served users, a GNN-based joint user scheduling and precoding method is investigated in \cite{he2022joint}. 
	
	Existing efforts in developing inherent properties in wireless communication systems are limited to GNN, requiring intricate node modeling and the construction of node update strategy during the design process. Therefore, with the increasing trend of incorporating multiple device types \cite{guo2021survey}, the design of schemes based on this approach may become increasingly complicated. Furthermore, although existing work has made contributions in developing equivariance in communication systems \cite{liu2023multidimensional, 9298921, wang2023soft}, there is little effort on investigating concise and unified frameworks to develop diverse equivariances such as multidimensional equivariance \cite{hartford2018deep}, higher-order equivariance \cite{keriven2019universal}, and invariance \cite{lee2019set} in such systems.
	
	In this paper, we focus on the development of these properties and proposed a unified framework for exploiting them in MU-MIMO systems. The major contributions of our work are summarized as follows:	
	
	\begin{itemize}
		\item \textbf{Framework Overview:} We propose the tensor equivariance (TE) framework for MU-MIMO transmission, which consists of four key components: \textit{mathematical foundation, module design, case study, and experimental validation}. These components collectively form a comprehensive solution that enables the efficient development of TE for key MU-MIMO communication problems.
		\item \textbf{Mathematical Foundation:} We explore the most valuable permutation equivariance/invariance properties and their variations (collectively referred to as TE) in MU-MIMO transmission through rigorous and concise mathematical definitions, thereby treating most past works as special cases of our approach. On the basis, we provide an alternative to topological modeling based on graphs or sets by introducing a Euclidean modeling approach that focuses on proving the inherent TE in mappings.
		This unifies and facilitates the development of TE under data-driven and model-driven methods, as well as supervised and unsupervised learning strategies.
		\item \textbf{Module Design:} Building on the mathematical foundations of TE, we design plug-and-play modules for each property. For multidimensional and high-order equivariance, we derived and provided equivalent models to existing methods, significantly simplifying deployment and enhancing scalability. For these modules, we further propose an approach that effectively decreases complexity while maintaining the network's ability to capture features across multiple dimensions. To model the invariance of the mapping across multiple dimensions, we construct a multidimensional invariant module based on a Transformer-based layer.
		\item \textbf{Case Study \& Experimental Validation:} 
		By integrating TE with the proposed plug-and-play modules, we streamline the design of TE networks for addressing MU-MIMO tasks. Based on the proposed framework, we focus on precoding and user scheduling, both key tasks in MU-MIMO transmission, as study cases. We not only design the corresponding networks, but also perform extensive experimental validation. The simulation results validate the effectiveness of all components of the TE framework by demonstrating its ability to reduce computational complexity and parameter count while maintaining strong generalizability across multiple dimensions. Specifically, in terms of user dimension, a network trained with 8 users can still generalize well and achieve satisfactory performance when applied to a scenario with 10 users.
	\end{itemize}

	This paper is structured as follows: In Section \ref{PE optimization problem sec}, we put forward TE in MU-MIMO systems. Section \ref{EIBs section} proposes plug-and-play TENN modules. Section \ref{NN design sec} investigates the unified TE framework. Section \ref{result sec} reports the simulation results, and the paper is concluded in Section \ref{conclusion}. 
	
	\textit{Standard Notation}: $(\cdot)^{-1}, (\cdot)^T, (\cdot)^H$ denote the inverse, transpose, and the transpose-conjugate operations, respectively. $x$, ${\bf x}$, ${\bf X}$, and $\tX$ respectively denote a scalar, column vector, matrix, and tensor. $\real(\cdot)$ and $\imaginary(\cdot)$ represent the real and imaginary part of a complex scalars, vector or matrix. $j=\sqrt{-1}$ denote imaginary unit. $\in$ denotes belonging to a set.  $\mathcal{A}\backslash\mathcal{B}$ means objects that belong to set $\mathcal{A}$ but not to $\mathcal{B}$. $|\mathcal{A}|$ represents the cardinality of set $\mathcal{A}$. ${\bf I}_{K}$ denotes $K\times K$ identity matrix. ${\bf 1}$ denotes the suitable-shape tensor with all elements being ones. $\left \|\cdot\right \|_{2}$ denotes $l_2$-norm. ${\rm det}({\bf A})$ represents the determinant of matrix ${\bf A}$. ${\rm blkdiag}\{{\bf A}_1,...,{\bf A}_K\}$ represents a block diagonal matrix composed of ${\bf A}_1,...,{\bf A}_K$. 
	
	\textit{Tensor Notation}: We use 
	$\tX_{[m_1,...,m_N]}$ to denote the indexing of elements in tensor $\tX\in \mathbb{R}^{M_1\times \cdots\times M_N}$. $[\tA_1,...,\tA_K]_{S}$ denotes the tensor formed by stacking $\tA_1,...,\tA_K$ along the $S$-th dimension. $[\cdot]_0$ denotes the concatenation of tensors along a new dimension, i.e., $\tA_{[n,:,\cdots]}=\tB_n,\ \forall n$ when $\tA=[\tB_1,...,\tB_N]_0$. We define the product of tensor $\tX\in \mathbb{R}^{M_1\times \cdots\times M_N\times D_X}$ and matrix ${\bf Y}\in \mathbb{R}^{D_X\times D_Y}$ as $(\tX\times {\bf Y})_{[m_1, ...,m_N,:]} = \tX_{[m_1, ...,m_N,:]}{\bf Y}$. The Kronecker product of tensor and matrix is defined as $(\tX\otimes_{n} {\bf Y})_{[m_1, ...,m_{n-1},:,:,m_{n+2},...,m_N]} = \tX_{[m_1, ...,m_{n-1},:,:,m_{n+2},...,m_N]}\otimes{\bf Y}$.
	
	\section{Tensor Equivariance in MU-MIMO systems}\label{PE optimization problem sec}
	
	\begin{figure*}[htp]
		\centering
		\includegraphics[width=7in]{Wang_Paper-TW-Jun-24-1235_fig1.eps}
		\caption{Some examples of TE. `Perm.' denotes the abbreviation of permutation; depth, pattern, and color are used to distinguish different features across the three dimensions; the solid-bordered arrows represent performing a certain permutation on the tensor, while the dashed-bordered arrows denote the permutation relations satisfied between tensors.}
		\label{fig TE}
	\end{figure*}
	
	In this section, we first present the concept of TE, and then propose a research paradigm for exploiting inherent TE within mappings of communication problems.
	
	\subsection{Tensor Equivariance}\label{TE subsection}
	Building on the properties in \cite{hartford2018deep} and \cite{maron2018invariant}, we propose the mathematical defintions of multidimensional equivariance and high-order equivariance, which offer improved generality and simplicity. Additionally, we extend the permutation invariance from \cite{zaheer2017deep} to multidimensional invariance, which captures the permutation invariance exhibited by mappings across multiple dimensions. We unify the mathematical expressions of these three properties, collectively referred to as TE. Below, we will provide their specific definitions. 
	
	The \textit{permutation} $\pi_M$ denotes a shuffling operation on the index $[1,...,M]$ of a length-$M$ vector under a specific pattern (or referred to as bijection from the the indices set ${\mathcal{M}} = \{1, 2, ..., M\}$ to ${\mathcal{M}}$), with the operator $\circ$ denoting its operation, and $\pi_M(m)$ represents the result of mapping $\pi_M$ on index $m$. For example, if $\pi_3 \circ [x_1, x_2, x_3]=[x_2, x_3, x_1]$, then $\pi_3(1) = 2$, $\pi_3(2) = 3$, and $\pi_3(3) = 1$ \cite{zaheer2017deep}. For tensor, we further extend the symbol $\circ$ to $\circ_n$, representing the permutation of dimension $n$ in the tensor by $\pi$. For instance, if $\pi_3 \circ_2 \tX=\tX'$, then $\tX_{[:, 1, :]} = \tX'_{[:, 3, :]}$, $\tX_{[:, 2, :]} = \tX'_{[:, 1, :]}$, and $\tX_{[:, 3, :]} = \tX'_{[:, 2, :]}$. We define the \textit{set of all permutations} for $[1,...,M]$ as ${\mathbb S}_M$, which is also referred to as the symmetric group \cite{artin2011algebra, ravanbakhsh2020universal}.  Then, we have $\pi_M\in{\mathbb S}_M$ and $|{\mathbb S}_M|=M!$.
	
	\figref{fig TE} illustrates several properties encompassed by TE in a diagrammatic form. Based on this, we will next introduce these properties.
	
	The mapping $f:\mathbb{R}^{M_1\times \cdots\times M_N\times D_X}\to \mathbb{R}^{M_1\times \cdots\times M_N\times D_Y}$ exhibits \textit{multidimensional ($N$-dimensional) equivariance} when it satisfies
	\begin{align}
		f({\pi_{M_n}}\!\circ_{n}\!\tX)={\pi_{M_n}}\!\circ_{n}\!f(\tX),\ \forall {\pi_{M_n}}\in {\mathbb{S}}_{M_n}, \forall n\in\mathcal{N}.
		\label{multidimensional equivariance}
	\end{align}
	where $\mathcal{N}=\{1,...,N\}$.
	This indicates that upon permuting a certain dimension in $\mathcal{N}$ of the input, the order of items in the corresponding dimension of the output will also be permuted accordingly \cite{yun2019transformers, kim2022pure}, which aligns with those described in \cite{hartford2018deep,maron2020learning}.
	
	We refer the mapping $f:\mathbb{R}^{\scriptsize{\overbrace{M\times \cdots\times M}^{p}}\times D_X}\to \mathbb{R}^{\scriptsize{\overbrace{M\times \cdots\times M}^{q}}\times D_Y}$ exhibits \textit{high-order ($p$-$q$-order) equivariance} when it satisfies
	\begin{align}
		f({\pi_{M}}\!\circ_{[1,...,p]}\!\tX)={\pi_{M}}\!\circ_{[1,...,q]}\!f(\tX),\ \forall {\pi_{M}}\in {\mathbb{S}}_{M},
		\label{high order layer}
	\end{align}
	where $\pi_{M}\circ_{[1,...,p]}$ represents performing the same permutation $\pi_{M}$ on the dimensions $1,...,p$, respectively. The equation expresses the equivariance of the mapping $f$ with respect to identical permutations across multiple dimensions, which originates from the descriptions in \cite{maron2018invariant, keriven2019universal}. We can distinguish between multidimensional and high-order equivariance by examining the permutation and the dimensions of the input and output. Specifically, for a mapping with the same input and output dimensions as $f$ in \eqref{high order layer}, it satisfies high-order equivariance whenever it satisfies multidimensional equivariance.
	
	The mapping $f:\mathbb{R}^{M_1\times \cdots\times M_N\times D_X}\to \mathbb{R}^{D_Y}$ exhibits \textit{multidimensional ($N$-dimensional) invariance} when \cite{lee2019set}
	\begin{align}
		f({\pi_{M_n}}\!\circ_{n}\!\tX)=f(\tX),\ \forall {\pi}_{M_n}\in {\mathbb{S}}_{M_n},\ \forall n\in\mathcal{N}.
		\label{multidimensional invariance}
	\end{align}
	The above equation illustrates that permuting the indices of the input across the dimensions contained in $\mathcal{N}$ does not affect the output of $f$. The properties described above are derived from the invariance in \cite{zaheer2017deep}.
	
	\subsection{Tensor Equivariance in Precoding Design}\label{precoding design sec}
	
	In Sections \ref{precoding design sec} and \ref{US design sec}, we develop a research paradigm for exploiting TE, whidch first construct the mapping between the available information and the optimal solution for the optimization problem, and then derive the TE that the mapping satisfies. As an example, we will validate the effectiveness of this paradigm for both data-driven and model-driven approaches in the precoding and user scheduling problems.
	
	Consider an MU-MIMO system where a BS equipped with $N_{\rm T}$ antennas transmits signals to $K$ users equipped with $N_{\rm R}$ antennas. 
	The optimization problem of sum-rate maximization can be formulated as \cite{6832894}:
	\begin{align}
		\begin{split}
			& \max_{\tW}\ \sum_{k=1}^KR_k(\tH,\tW,\sigma^2)\ \ {\rm s.t.}\ \sum_{k=1}^{K}{\rm Tr}\left({\bf W}_k{\bf W}^H_k\right)\leq P_{\rm T}, 
		\end{split}
		\label{3D precoding problem}
	\end{align} 
	where $\tH = [{\bf H}_1,...,{\bf H}_K]_{0}\in\mathbb{C}^{K\times N_{\rm R}\times N_{\rm T}}$, $\tW=[{\bf W}_1,...,{\bf W}_K]_0\in\mathbb{C}^{K\times N_{\rm R}\times N_{\rm T}}$, ${\bf H}_k\in{\mathbb{C}}^{N_{\rm R}\times N_{\rm T}}$ denotes the channel from the BS to the $k$-th user, ${\bf W}_k\in{\mathbb{C}}^{N_{\rm R}\times N_{\rm T}}$ denotes the precoding matrix of the $k$-th user, $P_{\rm T}$ represents the fixed transmit power, $\sigma^2$ is the noise power, $R_k(\tH,\tW,\sigma^2)=\log{\rm det}\left({\bf I}_k+{\bf W}_k{\bf H}^H_k{\boldsymbol {\Omega}}^{-1}_k{\bf H}_k{\bf W}^H_k\right)$ is the rate,
	and ${\boldsymbol{\Omega}}_k=\sigma^2{\bf I}+\sum_{i=1,i\neq k}^{K}{\bf H}_k{\bf W}^H_i{\bf W}_i{\bf H}^H_k\in{\mathbb{C}}^{N_{\rm R}\times N_{\rm R}}$ is the effective Interference-plus-noise covariance matrix.
	It can be concluded that \eqref{3D precoding problem} is a problem for $\tW$ based on available CSI $\tH$ and $\sigma^2$.  
	
	To simplify subsequent expressions, we define $\langle\tW,\{\tH,\sigma^2\}\rangle_{\rm P}$ as a pairing of precoding and CSI for problem \eqref{3D precoding problem}. Furthermore, the objective function achieved by $\tW$ and $\{\tH,\sigma^2\}$ in problem \eqref{3D precoding problem} is denoted as ``the objective function of $\langle\tW,\{\tH,\sigma^2\}\rangle_{\rm P}$''.
	On this basis, the property of optimization problem \eqref{3D precoding problem} is as follows.
	
	\begin{ppn}\label{ppn precoding}
		The objective function of  $\langle\tW,\{\tH,\sigma^2\}\rangle_{\rm P}$ is equal to those of $\langle\pi_{K}\circ_1\tW,\{\pi_{K}\circ_1\tH,\sigma^2\}\rangle_{\rm P}$, $\langle\pi_{N_{\rm R}}\circ_2\tW,\{\pi_{N_{\rm R}}\circ_2\tH,\sigma^2\}\rangle_{\rm P}$, and $\langle\pi_{{N_{\rm T}}}\circ_3\tW,\{\pi_{{N_{\rm T}}}\circ_3\tH,\sigma^2\}\rangle_{\rm P}$, for all $\pi_{K}\in{\mathbb{S}}_K$, $\pi_{N_{\rm R}}\in{\mathbb{S}}_{N_{\rm R}}$, and $\pi_{{N_{\rm T}}}\in{\mathbb{S}}_{N_{\rm T}}$. Specifically, if $\langle\tW^{\star},\{\tH,\sigma^2\}\rangle_{\rm P}$ achieves the optimal objective function, then $\langle\pi_{K}\circ_1\tW^{\star},\{\pi_{K}\circ_1\tH,\sigma^2\}\rangle_{\rm P}$, $\langle\pi_{N_{\rm R}}\circ_2\tW^{\star},\{\pi_{N_{\rm R}}\circ_2\tH,\sigma^2\}\rangle_{\rm P}$, and $\langle\pi_{{N_{\rm T}}}\circ_3\tW^{\star},\{\pi_{{N_{\rm T}}}\circ_3\tH,\sigma^2\}\rangle_{\rm P}$ can also achieve their optimal objective functions.
		\label{precoding 3D PE ppn}
	\end{ppn}
	\begin{pf}
		See Appendix \ref{pf precoding}.
	\end{pf}

	More clearly, we define $G_{\rm P}(\cdot)$ as a mapping from available CSI to one of the optimal precoding schemes for \eqref{3D precoding problem}, i.e., $G_{\rm P}(\tH, \sigma^2)=\tW^{\star}$. 
	Then, based on \ppnref{ppn precoding}, the following equations hold when problem \eqref{3D precoding problem} has a unique optimal solution.
	\begin{subequations}
		\begin{align}
			&G_{\rm P}({\pi_K}\!\circ_{1}\!\tH, \sigma^2)={\pi_K}\!\circ_{1}\!\tW^{\star},\ \forall {\pi}_K\in {\mathbb{S}}_K,\label{ppn precoding e1}\\
			&G_{\rm P}({\pi_{N_{\rm R}}}\!\circ_{2}\!\tH, \sigma^2)={\pi_{N_{\rm R}}}\!\circ_{2}\!\tW^{\star},\ \forall {\pi}_{N_{\rm R}}\in {\mathbb{S}}_{N_{\rm R}},\label{ppn precoding e2}\\
			&G_{\rm P}({\pi_{N_{\rm T}}}\!\circ_{3}\!\tH, \sigma^2)={\pi_{N_{\rm T}}}\!\circ_{3}\!\tW^{\star},\ \forall {\pi}_{N_{\rm T}}\in {\mathbb{S}}_{N_{\rm T}}.\label{ppn precoding e3}
		\end{align}
	\end{subequations}
	
	If the optimization problem has multiple optimal solutions, it can be proven that the problem based on the permuted CSI also has the same number of optimal solutions, and the optimal solutions of the two problems correspond one-to-one in the manner described by the equations above. In this case, $G_{\rm P}(\cdot)$ can be regarded as a mapping to certain single optimal solution. To avoid potential performance degradation caused by fitting multiple mappings simultaneously, we can feed the initial solution as additional input to the network \cite{sun2018learning}.
	
	Then, we consider the TE in precoding design under the model-driven approach. For problem \eqref{3D precoding problem}, iterative algorithms based on optimal closed-form expressions can achieve outstanding performance and have garnered considerable attention \cite{4712693,6832894}.
	The well-known solution to problem \eqref{3D precoding problem} can be obtained through the following expression \cite{4712693,shi2011iteratively, zhao2023rethinking}
	\begin{gather}
		\begin{split}
			{\bf W} \!=\! \gamma{\tilde{\bf W}},{\tilde{\bf W}} \!=\! {\bf H}^H{\bf A}^H{\bf U}\left(\mu{\bf I}_{MK}\!+\!{\bf A}{\bf H}{\bf H}^H{\bf A}^H{\bf U}\right)^{-1},\\
			\gamma =\sqrt{{P_{\rm T}}/{{\rm Tr}({\tilde{\bf W}}{\tilde{\bf W}}^H)}},\  \mu = {{\rm Tr}\left({\bf U}{\bf A}{\bf A}^H\right)}\sigma^2/{P_{\rm T}},\label{precoding CF}
		\end{split}
	\end{gather}
	where 
	\begin{gather}
		{\bf W} = [{\bf W}_1,...,{\bf W}_K]^T\in{\mathbb C}^{N_{\rm T}\times KN_{\rm R}},\\
		{\bf U} = {\rm blkdiag}\{{\bf U}_1,...,{\bf U}_K\}\in{\mathbb C}^{KN_{\rm R}\times KN_{\rm R}},\\
		{\bf A} = {\rm blkdiag}\{{\bf A}_1,...,{\bf A}_K\}\in{\mathbb C}^{KN_{\rm R}\times KN_{\rm R}},\\
		{\bf H} = [{\bf H}^T_1,...,{\bf H}^T_K]^T\in{\mathbb C}^{KN_{\rm R}\times N_{\rm T}},
	\end{gather}
	where ${\bf U}_k\in{\mathbb{C}}^{N_{\rm R}\times N_{\rm R}}$ and the Hermitian matrix ${\bf A}_k\in{\mathbb{C}}^{N_{\rm R}\times N_{\rm R}}$ are auxiliary tensors that require iterative computations with relatively high computational complexity to obtain based on $\{\tH,\sigma^2\}$ \cite{4712693}. To simplify the expression, we represent the aforementioned closed-form computation as $\tW = {\rm CFP}\left(\tH,\tA,\tU,\sigma^2\right)$, where $\tA=[{\bf A}_1,...,{\bf A}_K]_{0}\in\mathbb{C}^{K\times N_{\rm R}\times N_{\rm R}}$ and $\tU=[{\bf U}_1,...,{\bf U}_K]_{0}\in\mathbb{C}^{K\times N_{\rm R}\times N_{\rm R}}$.
	
	We define $\langle\{\tA, \tU\},\{\tH,\sigma^2\}\rangle_{\rm CFP}$ as a pairing of auxiliary tensors and CSI for the closed-form expression to problem \eqref{3D precoding problem}. The objective function achieved by $\tW = {\rm CFP}\left(\tH,\tA,\tU,\sigma^2\right)$ and $\{\tH,\sigma^2\}$ in problem \eqref{3D precoding problem} is denoted by ``the objective function of $\langle\{\tA, \tU\},\{\tH,\sigma^2\}\rangle_{\rm CFP}$''.
	
	\begin{ppn}\label{ppn CF precoding}
		The objective function of $\langle\{\tA, \tU\},\{\tH,\sigma^2\}\rangle_{\rm CFP}$ is equal to those of $\langle\{\pi_{K}\circ_1\! \tA, \pi_{K}\circ_1\!\tU\},\{\pi_{K}\circ_1\!\tH,\sigma^2\}\rangle_{\rm CFP}$, $\langle\{\pi_{N_{\rm R}}\circ_{[2,3]}\!\tA, \pi_{N_{\rm R}}\circ_{[2,3]}\!\tU\},\{\pi_{N_{\rm R}}\circ_2\!\tH,\sigma^2\}\rangle_{\rm CFP}$, and $\langle\{\tA, \tU\},\{\pi_{N_{\rm T}}\circ_3\!\tH,\sigma^2\}\rangle_{\rm CFP}$, for all $\pi_{K}\in{\mathbb{S}}_K$, $\pi_{N_{\rm R}}\in{\mathbb{S}}_{N_{\rm R}}$, and $\pi_{{N_{\rm T}}}\in{\mathbb{S}}_{N_{\rm T}}$. Specifically, if $\langle\{\tA, \tU\},\{\tH,\sigma^2\}\rangle_{\rm CFP}$ achieves the optimal objective function\footnote{The optimal objective function here refers to the maximum achievable objective function of the closed-form expression in \eqref{precoding CF}.}, then $\langle\{\pi_{K}\circ_1\!\tA^\star, \pi_{K}\circ_1\!\tU^\star\},\{\pi_{K}\circ_1\!\tH,\sigma^2\}\rangle_{\rm CFP}$, $\langle\{\pi_{N_{\rm R}}\circ_{[2,3]}\!\tA^\star, \pi_{N_{\rm R}}\circ_{[2,3]}\!\tU^\star\},\{\pi_{N_{\rm R}}\circ_2\!\tH,\sigma^2\}\rangle_{\rm CFP}$, and $\langle\{\tA^\star, \tU^\star\},\{\pi_{N_{\rm T}}\circ_3\!\tH,\sigma^2\}\rangle_{\rm CFP}$ can also achieve their optimal objective functions.
	\end{ppn}
	\begin{pf}
		See Appendix \ref{pf CF precoding}.
	\end{pf}
	
	Similar to \ppnref{precoding 3D PE ppn}, we define $G_{\rm CFP}(\cdot)$ as a mapping from available CSI to one pair of the optimal auxiliary tensors for the closed-form expression \eqref{precoding CF} of problem \eqref{3D precoding problem}, i.e., $G_{\rm CFP}(\tH, \sigma^2)=\tA^\star, \tU^\star$. When problem \eqref{3D precoding problem}'s closed-form \eqref{precoding CF} has only one pair of optimal auxiliary tensors, for all ${\pi}_K$, ${\pi}_{N_{\rm R}}$, and ${\pi}_{N_{\rm T}}$ belonging to ${\mathbb{S}}_K$, ${\mathbb{S}}_{N_{\rm R}}$, and ${\mathbb{S}}_{N_{\rm T}}$, respectively, the following equations hold based on \ppnref{ppn CF precoding}.
	\begin{subequations}
		\begin{align}
			&G_{\rm CFP}({\pi_K}\!\circ_{1}\!\tH, \sigma^2)=\pi_{K}\circ_1\tA^\star, \pi_{K}\circ_1\tU^\star, \label{ppn CF precoding e1}\\
			&G_{\rm CFP}({\pi_{N_{\rm R}}}\!\circ_{2}\!\tH, \sigma^2)=\pi_{N_{\rm R}}\circ_{[2,3]}\tA^\star, \pi_{N_{\rm R}}\circ_{[2,3]}\tU^\star, \label{ppn CF precoding e2}\\
			&G_{\rm CFP}({\pi_{N_{\rm T}}}\!\circ_{3}\!\tH, \sigma^2)=\tA^\star, \tU^\star. \label{ppn CF precoding e3}
		\end{align}
	\end{subequations}
	
	Furthermore, for MU-MISO systems, the closed-form expression in \eqref{precoding CF} will degenerate to the closed-form expression in \cite{6832894}. Except for the aspects related to the permutation of receive antennas, the remaining content in \ppnref{ppn CF precoding} remains valid. Similar properties are applicable to the preprocessed CSI and the modified closed-form in \cite{zhang2022deep}.
	
	\subsection{Tensor Equivariance in User Scheduling Design}\label{US design sec}
	In this subsection, we consider the design of downlink user scheduling for the system in Section \ref{precoding design sec}, where $K$ users are selected from ${\tilde K}$ candidate users for downlink transmission, and the $K$ users utilize a certain precoding scheme $\tW = G_{\rm CP}(\tH, \sigma^2)$ for the downlink transmission. Without loss of generality, we assume that $G_{\rm CP}(\cdot)$ possesses the properties described by \eqref{ppn precoding e1}-\eqref{ppn precoding e3}. The user scheduling problem for sum-rate maximization is given by
	\begin{align}
		\begin{split}
			&\max_{{\boldsymbol{\eta}}}\ R_{\rm US}({\tilde \tH},{\boldsymbol{\eta}},\sigma^2)\\
			&\qquad{\rm s.t.}\ \eta_{{\tilde k}}\in\{0,1\},\ {\tilde k}\in{\tilde{\mathcal{K}}}, \\
			&\qquad\quad\  \textstyle{\sum_{{\tilde k}\in {\tilde {\mathcal{K}}}}\eta_{\tilde k} = K},
		\end{split}
		\label{user scheduling problem}
	\end{align}
	where
	\begin{gather}
		R_{\rm US}({\tilde \tH},{\boldsymbol{\eta}},\sigma^2) = \sum_{k\in {\mathcal{K}}}R_k(\tH,G_{\rm CP}(\tH, \sigma^2),\sigma^2),\\
		{\mathcal{K}}\!=\!\{{k}|\eta_{k}\!=\!1,{k}\!\in\!{{\tilde{\mathcal{K}}}}\}, \ {\tilde \tH} = [{\bf H}_k]_{0,k\in{\tilde{\mathcal{K}}}}\in\mathbb{C}^{{\tilde K}\times N_{\rm R}\times N_{\rm T}},
	\end{gather}
	$\eta_{{\tilde k}}$ is the scheduling indicator for user ${\tilde k}$, and ${\boldsymbol{\eta}}=[\eta_1,...,\eta_{\tilde K}]^T\in\mathbb{C}^{{\tilde K}\times 1}$. \eqref{user scheduling problem} is a problem for ${\boldsymbol{\eta}}$ based on ${\tilde {\tH}}$ and $\sigma^2$.  
	
	Similar to Section \ref{precoding design sec}, we define $\langle{\boldsymbol{\eta}},\{{\tilde {\tH}},\sigma^2\}\rangle_{\rm US}$ for problem \eqref{user scheduling problem}, and the property of this problem is as follows.
	
	\begin{ppn}\label{ppn US}
		The objective function of $\langle{\boldsymbol{\eta}},\{{\tilde {\tH}},\sigma^2\}\rangle_{\rm US}$ is equal to those of $\langle\pi_{{\tilde K}}\circ_1{\boldsymbol{\eta}},\{\pi_{{\tilde K}}\circ_1{\tilde {\tH}},\sigma^2\}\rangle_{\rm US}$, $\langle{\boldsymbol{\eta}},\{\pi_{N_{\rm R}}\circ_2{\tilde {\tH}},\sigma^2\}\rangle_{\rm US}$, and $\langle{\boldsymbol{\eta}},\{\pi_{{N_{\rm T}}}\circ_3{\tilde {\tH}},\sigma^2\}\rangle_{\rm US}$, for all $\pi_{\tilde K}\in{\mathbb{S}}_{\tilde K}$, $\pi_{N_{\rm R}}\in{\mathbb{S}}_{N_{\rm R}}$, and $\pi_{{N_{\rm T}}}\in{\mathbb{S}}_{N_{\rm T}}$. Furthermore, if $\langle{\boldsymbol{\eta}}^{\star},\{{\tilde {\tH}},\sigma^2\}\rangle_{\rm US}$ achieves the optimal objective function, then $\langle\pi_{{\tilde K}}\circ_1{\boldsymbol{\eta}}^{\star},\{\pi_{{\tilde K}}\circ_1{\tilde {\tH}},\sigma^2\}\rangle_{\rm US}$, $\langle{\boldsymbol{\eta}}^{\star},\{\pi_{N_{\rm R}}\circ_2{\tilde {\tH}},\sigma^2\}\rangle_{\rm US}$, and $\langle{\boldsymbol{\eta}}^{\star},\{\pi_{{N_{\rm T}}}\circ_3{\tilde {\tH}},\sigma^2\}\rangle_{\rm US}$ can also achieve their optimal objective functions.
	\end{ppn}
	
	\begin{pf}
		See Appendix \ref{pf US}.
	\end{pf}
	
	We define $G_{\rm US}(\cdot)$ as a mapping from available CSI to one of the optimal binary selection variables for \eqref{user scheduling problem}, i.e., $G_{\rm US}({\tilde {\tH}}, \sigma^2)={\boldsymbol{\eta}}^{\star}$. 
	Then, based on \ppnref{ppn US}, the following equations hold when problem \eqref{user scheduling problem} has a unique optimal solution.
	\begin{subequations}
		\begin{align}
			&G_{\rm US}({\pi_{\tilde K}}\!\circ_{1}\!{\tilde {\tH}}, \sigma^2)={\pi_{\tilde K}}\!\circ_{1}\!{\boldsymbol{\eta}}^{\star},\ \forall {\pi}_{\tilde K}\in {\mathbb{S}}_{\tilde K},\label{ppn US e1}\\
			&G_{\rm US}({\pi_{N_{\rm R}}}\!\circ_{2}\!{\tilde {\tH}}, \sigma^2)={\boldsymbol{\eta}}^{\star},\ \forall {\pi}_{N_{\rm R}}\in {\mathbb{S}}_{N_{\rm R}},\label{ppn US e2}\\
			&G_{\rm US}({\pi_{N_{\rm T}}}\!\circ_{3}\!{\tilde {\tH}}, \sigma^2)={\boldsymbol{\eta}}^{\star},\ \forall {\pi}_{N_{\rm T}}\in {\mathbb{S}}_{N_{\rm T}}.
			\label{ppn US e3}
		\end{align}
	\end{subequations}
	
	\section{Tensor Equivariance NN Modules}\label{EIBs section}
	
	In the previous section, we revealed the multidimensional equivariance (such as \eqref{ppn precoding e1}-\eqref{ppn precoding e3}), high-order equivariance (such as \eqref{ppn CF precoding e2}), and invariance (such as \eqref{ppn CF precoding e3}, \eqref{ppn US e2}, and \eqref{ppn US e3}) in the mappings required for MU-MIMO systems. In this section, we develop plug-and-play TENN modules that satisfy these properties, thereby laying the groundwork for constructing NNs for approximating mappings in MU-MIMO systems. 
	
	\subsection{Multi-Dimensional Equivariant Module}\label{PE linear layer sec}
	In this subsection, we investigate function satisfying multidimensional equivariance to approximate the mapping like those in \eqref{ppn precoding e1}-\eqref{ppn precoding e3}. The conventional FC layer processing involves flattening the features, multiplying them with a weight matrix, and adding bias. The operation ${\rm FC}(\cdot):\mathbb{R}^{M_1\times\cdots\times M_N\times D_{\rm I}} \rightarrow \mathbb{R}^{M_1\times\cdots\times M_N\times D_{\rm O}}$ is as follows
	\begin{align}
		\tY = {\rm FC}(\tX)= {\rm vec}^{-1}\left({\bf W}{\rm vec}(\tX)+{\bf b}\right),
		\label{conventional linear layer}
	\end{align}
	where $\tX\in\mathbb{R}^{M_1\times\cdots\times M_N\times D_{\rm I}}$ denotes the input, $\tY\in\mathbb{R}^{M_1\times\cdots\times M_N\times D_{\rm O}}$ represents the output, ${\bf W}\in\mathbb{R}^{{\bar M}D_{\rm I}\times {\bar M}D_{\rm O}}$ denotes the weight, ${\bf b}\in\mathbb{R}^{{\bar M}D_{\rm O}\times 1}$ denotes the bias, and ${\bar M}=\Pi^{N}_{n=1}M_n$. We refer $D_{\rm I}$ and $D_{\rm O}$ as the feature lengths. 
	
	The necessary and sufficient expressions of ${\bf W}$ in layer $\tY = {\rm vec}^{-1}\left({\bf W}{\rm vec}(\tX)\right)$ with $N$-dimensional equivariance is demonstrated in \cite{hartford2018deep}. However, the intricate patterns of ${\bf W}$ make its construction and deployment challenging, and it becomes increasingly difficult as the dimensionality increases. To this end, we derive a new equivalent mathematical expression that involves only simple operations, such as averaging across dimensions, making it easy to implement. Furthermore, we also consider the derivation of the patterns of ${\bf b}$. Based on this, we propose a method to significantly reduce the complexity of this module while maintaining nearly the same performance.
	
\begin{figure}[htbp]
		\centering
		\includegraphics[width=0.48\textwidth]{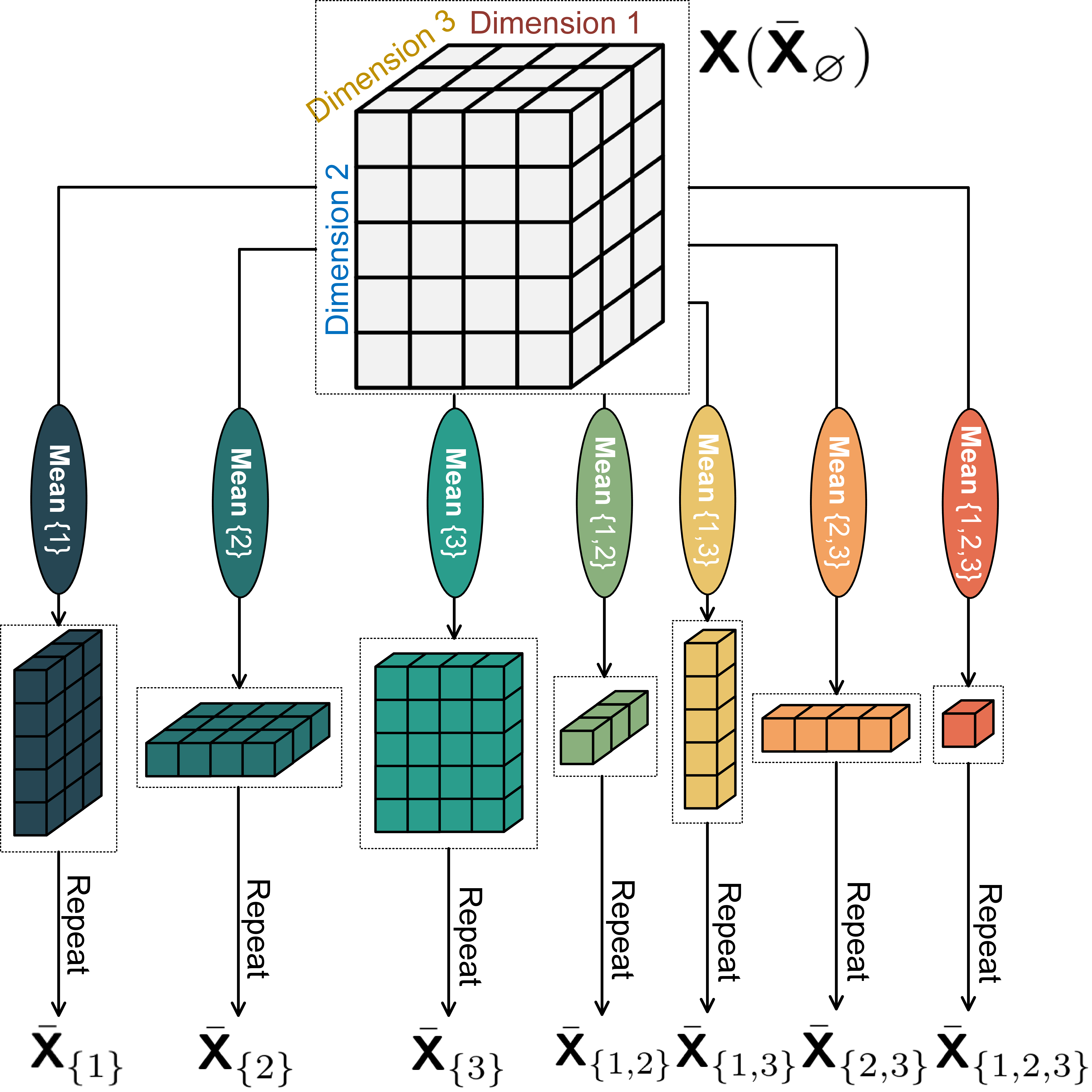}
		\caption{In the case of $N=3$, the acquisition for all ${\bar \tX}_{\mathcal P},\ {\mathcal P}\subseteq {\mathcal N}$.}
		\label{MDPE_Mean}
	\end{figure}

	Firstly, to simplify subsequent expressions, we define ${\bar \tX}_{\mathcal P}$ to represent the result of averaging $\tX$ along the dimensions in ${\mathcal{P}}$ and then repeating it to the original dimensions. Note that ${\bar \tX}_{\varnothing}=\tX$. Besides, we denote ${\bar {\mathcal N}}$ as the set containing all subsets of ${\mathcal N}=\{1,..,N\}$ (power set). As an example, when $N=3$, \figref{MDPE_Mean} illustrates the acquisition of all ${\bar \tX}_{\mathcal P},\ {\mathcal P}\in{\bar {\mathcal N}}$. 
	
	\begin{ppn}\label{ppn linear PE}
		Any FC layer $\tY = {\rm FC}(\tX)$ satisfying multidimensional equivariance across dimensions $M_1, M_2,...,M_N$ can be represented as
		\begin{align}
			{\rm FC}_{\rm MDE}(\tX) = \sum_{{\mathcal P}\in {\bar {\mathcal N}}}\left({\bar \tX}_{\mathcal P}\times\!{\bf W}_{{\mathcal{P}}}\right)+{\bf 1}\!\otimes_N\!{\bf b}^T_{\rm MDE},
			\label{PE linear layer}
		\end{align}
		where ${\bf W}_{{\mathcal{P}}}\in\mathbb{R}^{D_{\rm I}\times D_{\rm O}},\ \forall {\mathcal P}\in {\bar {\mathcal N}}$ and ${\bf b}_{\rm MDE}\in\mathbb{R}^{D_{\rm O}\times 1}$ are learnable parameters.
	\end{ppn}
	\begin{pf}
		See Appendix \ref{pf linear PE}.
	\end{pf}
	\ppnref{ppn linear PE} indicates that when equivariance is satisfied across $N$ dimensions, the processing of the FC layer degenerates into linear combination of the means of the input tensor at all dimension combinations in ${\mathcal N}$. By defining ${\rm FFC}(\cdot)$ as the learnable FC layer that applied to the last dimension, i.e., ${\rm FFC}(\tX) = \tX\times {\bf W}+{\bf 1}\!\otimes_N\!{\bf b}$, the layer in \eqref{PE linear layer} can be achieved by ${\rm FC}_{\rm {MDE}}(\tX)={\rm FFC}([{\bar \tX}_{\mathcal P}]_{N+1,{\mathcal P}\in {\bar {\mathcal N}}})$.
	
	Then, we analyze the changes in computational complexity and the number of parameters brought about by this degeneration. In \eqref{conventional linear layer}, for ${\rm FC}(\cdot)$, the number of multiplications is $\mathcal{O}({\bar M}^2D_{\rm I}D_{\rm O})$, and the number of parameters is $\mathcal{O}({\bar M}^2D_{\rm I}D_{\rm O})$. For ${\rm FC}_{\rm {MDE}}(\cdot)$, since the power set of ${\mathcal N}$ consists of $2^N$ elements, the computational complexity and the number of parameters are ${\mathcal O}(2^N{\bar M}D_{\rm I}D_{\rm O})$ and ${\mathcal O}(2^ND_{\rm I}D_{\rm O})$, respectively. Given ${\bar M} = \Pi^{N}_{n=1}M_n$, it holds that ${\bar M}\gg 2^N$ when $M_n>2, n\in{\mathcal N}$, which means ${\rm FC}_{\rm {MDE}}(\cdot)$ significantly reduces the complexity. Note that the computational complexity of ${\rm FC}(\cdot)$ is determined by ${\bar M}^2$, while that of ${\rm FC}_{\rm {MDE}}(\cdot)$ is only determined by ${\bar M}$. Furthermore, the number of parameters in ${\rm FC}(\cdot)$ is dependent on ${\bar M}$, while that of ${\rm FC}_{\rm {MDE}}(\cdot)$ is solely determined by $N$. 
	
	The computational complexity and parameter count of the module are related to $N$ by the relationship $\mathcal{O}(|{\bar {\mathcal N}}|)=\mathcal{O}(2^N)$, and the complexity becomes high when $N$ is large. To this end, we propose the approach to reduce the complexity, whose main idea is to replace the set ${\bar {\mathcal N}}$ by its special subset ${\bar {\mathcal N}}'$. 
	$\tX_{{\mathcal P}}$ represents one-dimensional and cross-dimensional global features in the case of $|{\mathcal P}|=1$ and $|{\mathcal P}|>1$, respectively. For a single-layer network, if we remove certain ${\mathcal P}$ from the set ${\bar {\mathcal N}}$ (where ${\mathcal P}\neq \varnothing$), the network will lose the ability to capture global features in those dimension combinations. Fortunately,
	for a multi-layer network, even if some ${\mathcal P}$ are removed, the network may still retain the capability to capture global features of various dimensional combinations through multi-layer stacking, similar to the receptive field mechanism in CNN \cite{luo2016understanding}. Based on this, we propose the following three patterns for constructing set ${\bar {\mathcal N}}'$, where ${\bar {\mathcal N}}'_l$ represents the set of the $l$-th layer.
	\begin{itemize}
		\item \textbf{Pattern 1:} The set ${\bar {\mathcal N}}'_l$ for each layer of the network is the same, which is 
		\begin{align}
			{\bar {\mathcal N}}^{\rm P1}_{l}=\{\varnothing, \{1\},\{2\},...,\{N\}\},\ \forall l.
		\end{align}
		When $N=3$, we have ${\bar {\mathcal N}}^{\rm P1}_{l}=\{\varnothing, \{1\},\{2\},\{3\}\}$. The complexity order is reduced from ${\mathcal O}(2^N)$ to ${\mathcal O}(N+1)$.
		\item \textbf{Pattern 2:} The set ${\bar {\mathcal N}}'_{l}$ for each layer of the network can be different, but must include the empty set $\varnothing$. The included sets must satisfy $|{\mathcal P}|\leq 1$ and need to satisfy
		\begin{align}
			{\bar {\mathcal N}}^{\rm P2}_{1}\cup {\bar {\mathcal N}}^{\rm P2}_{2}\cup\cdots \cup{\bar {\mathcal N}}^{\rm P2}_{L} = \{\varnothing, \{1\},...,\{N\}\}.
			\label{pattern condition}
		\end{align}
		For example, when $N=3$ and $L=3$, the sets can be ${\bar {\mathcal N}}^{\rm P2}_{1} = \{\varnothing, \{1\}, \{2\}\}$,  ${\bar {\mathcal N}}^{\rm P2}_{2} = \{\varnothing, \{2\}, \{3\}\}$, and  ${\bar {\mathcal N}}^{\rm P2}_{3} = \{\varnothing, \{1\}, \{3\}\}$. The above equation ensures that there is no loss of global feature for all dimensions throughout the entire network. The complexity order is reduced from ${\mathcal O}(2^N)$ to ${\mathcal O}(N')$, where $N'\leq N+1$.
		\item \textbf{Pattern 3:} The sets ${\bar {\mathcal N}}'_{l}$ of different layers can be different, but they only contain two elements, $\varnothing$ and ${\mathcal P}_l$, where $|{\mathcal P}_l|=1$, and they need to satisfy
		\begin{align}
			{\mathcal P}_{1}\cup {\mathcal P}_{2}\cup\cdots \cup{\mathcal P}_{L} = \{\{1\},...,\{N\}\}.
		\end{align}
		For example, when $N=3$ and $L=3$, the sets can be ${\bar {\mathcal N}}^{\rm P3}_{1} = \{\varnothing, \{1\}\}$,  ${\bar {\mathcal N}}^{\rm P3}_{2} = \{\varnothing, \{2\}\}$, and  ${\bar {\mathcal N}}^{\rm P3}_{3} = \{\varnothing, \{3\}\}$. This pattern can be regarded as a special case of \textbf{Pattern 2}.
		The complexity order is reduced from ${\mathcal O}(2^N)$ to ${\mathcal O}(2)$.
	\end{itemize}
	
	\subsection{High-Order Equivariant Module}\label{PE HighOd layer sec}
	In this subsection, we construct functions satisfying high-order equivariance.
	The necessary and sufficient parameter-sharing pattern for ${\bf W}$ and ${\bf b}$ in layer $\tY = {\rm vec}^{-1}\left({\bf W}{\rm vec}(\tX) + {\bf b}\right)$ with $p$-$q$-order equivariance is demonstrated in \cite{maron2018invariant}, which describes the element-wise relationship between the output and input under complex set definitions. Based on this work, we propose a new mathematically equivalent model that is easier to understand, construct, and deploy.
	
	\begin{ppn}\label{ppn linear highOd PE}
		Any FC layer $\tY = {\rm FC}(\tX)$ satisfying the $p$-$q$-order equivariance can be represented as
		\begin{align}
			{\rm FC}_{\rm HOE}(\tX) = \sum_{{\mathcal P}\in{\tilde {\mathcal{N}}}_{p\!-\!q}}\left({\tilde \tX}_{\mathcal P}\times\!{\bf W}_{{\mathcal P}}\right)+\sum_{{\mathcal Q}\in{\tilde {\mathcal{N}}}_{q}}\tB_{{\mathcal Q}}
			\label{highOd PE linear layer}
		\end{align}
		where ${\bf W}_{{\mathcal P}}\in\mathbb{R}^{D_{\rm I}\times D_{\rm O}}$ is the learnable matrix for ${\tilde \tX}_{\mathcal P}$, and $\tB_{{\mathcal Q}}$ denotes the bias constructed based on set ${\mathcal Q}$. ${\tilde \tX}_{\mathcal P}$ represents the output tensor formed by the simple combination of elements from the input tensor based on set ${\mathcal P}$. The construction of ${\tilde {\mathcal{N}}}$, ${\tilde \tX}_{\mathcal P}$, and $\tB_{{\mathcal Q}}$ can be found in Appendix \ref{HOE detail}.
	\end{ppn}
	
	The proposition can be proved based on \cite{maron2018invariant} and \cite{pan2022permutation}. Similar to ${\rm FC}_{\rm MDE}(\cdot)$, ${\rm FC}_{\rm HOE}(\cdot)$ essentially performs a linear combination of the matrices construted by input itself and its global feature across each dimension and their combinations.
	The computational complexity and parameter count of ${\rm FC}(\cdot)$ are both ${\mathcal O}(M^{p+q}D_{\rm I}D_{\rm O})$, while those of ${\rm FC}_{\rm HOE}(\cdot)$ are ${\mathcal O}(|{\tilde {\mathcal{N}}}_{p\!-\!q}|M^qD_{\rm I}D_{\rm O})$ and ${\mathcal O}(D_{\rm I}D_{\rm O})$, respectively. Note that $|{\tilde {\mathcal{N}}}_{p\!-\!q}| = B(p+q)$, where $B(\cdot)$ is the Bell number. Specifically, the construction details of the 1-2-order equivariant module are provided in Appendix \ref{HOE detail}. When applied to tensors, one can only do the operation for certain two dimensions and regard the other dimensions as batch dimensions. For example, when we apply 1-2-order ${\rm FC}_{\rm HOE}(\cdot):\mathbb{R}^{M\times D_{\rm I}}\to \mathbb{R}^{M\times M\times D_{\rm O}}$ to the last two dimensions of $\tX\in\mathbb{R}^{M_1\times\cdots\times M_N\times D_{\rm I}}$, for all $m_1\in\{1,...,M_1\}$,..., $m_{N-1}\in\{1,...,M_{N-1}\}$, execute the following same operation
	\begin{align}
		{\tY}_{[m_1,...,m_{N\!-\!1},:,:,:]} \!=\! {\rm FC}_{\rm HOE}(\tX_{[m_1,...,m_{N\!-\!1},:,:]})
	\end{align}
	where ${\tY}_{[m_1,...,m_{N\!-\!1},:,:,:]}\in \mathbb{R}^{M_N\!\times\! M_N\!\times\! D_{\rm O}}$. The above operation can be executed in parallel through batch computation. 
	
	It is worth noting that the mathematical expression of \eqref{highOd PE linear layer} is consistent with the multidimensional equivariant module \eqref{PE linear layer}, which means that we can adopt a similar approach by constructing subset ${\tilde {\mathcal{N}}}'_{p\!-\!q,l}$ to replace set ${\tilde {\mathcal{N}}}_{p\!-\!q}$, thereby significantly reducing the complexity. Specifically, when constructing a network by stacking multiple high-order equivariant modules, we can either remove all ${\mathcal P}\in {\tilde {\mathcal{N}}}_{p\!-\!q}$ containing features across multiple dimensions (Pattern 1); or, further refine each module by removing parts of the ${\mathcal P}\in {\tilde {\mathcal{N}}}_{p\!-\!q}$ that contain features within every single dimension (Pattern 2); or construct the set ${\tilde {\mathcal{N}}}'_{p\!-\!q,l}$ for each module by retaining only ${\mathcal P}$ that keeps specific single-dimensional features (Pattern 3).
	
	\subsection{Multi-Dimensional Invariant Module}\label{IE Attention layer sec}
	In this subsection, we first introduce function pooling by multihead attention (PMA) in \cite{lee2019set}, which satisfies single-dimensional invariance. On this basis, we extend PMA and propose the multidimensional invariant module, which can capture the invariance of mappings across multiple dimensions. The expression of PMA is as follows
	\begin{align}
		{\rm PMA}({\bf X}) = {\rm MAB}({\bf S}, {\rm FFC}({\bf X})) \in\mathbb{R}^{J\times D_{\rm O}}, 
		\label{PMA eq1}
	\end{align}
	where ${\bf X}\in\mathbb{R}^{M\times D_{\rm I}}$ is the input matrix, and ${\bf S}\in\mathbb{R}^{J\times D_{\rm I}}$ is a learnable parameter matrix, $M$ denotes the number of input items, and $D_{\rm O}$ represents the output feature length. $J$ controls the dimension of the output matrix. Without loss of generality, we all subsequently set $J=1$. The expression of multi-head attention block (MAB) is given by\footnote{The matrices in the expression are only used for illustrative purposes, so the matrix dimensions are not given.}
	\begin{gather}
		{\rm MAB}({\bf X}', {\bf Y}') = {\bf M}'+{\rm ReLU}\left({\rm FFC}({\bf M}')\right),\label{MAB eq1}\\
		{\bf M}' = {\rm LN}({\bf X}'+{\rm MultiHead}({\bf X}', {\bf Y}', {\bf Y}')).
	\end{gather}
	${\rm MultiHead}(\cdot)$ denotes the multi-head attention module \cite{vaswani2017attention}, whose expression can be written as ${\rm MultiHead}({\bf Q}, {\bf K}, {\bf V}) = [{\rm head}_1,...,{\rm head}_{N_{\rm H}}]{\bf W}^{O}$, where
	\begin{gather}
		{\rm head}_i = {\rm Attention}({\bf Q}{\bf W}^{Q}_i, {\bf K}{\bf W}^{K}_i, {\bf V}{\bf W}^{V}_i),
	\end{gather}
	where ${\bf W}^{Q}, {\bf W}^{K}, {\bf W}^{V}\in{\mathbb R}^{D_{\rm I}\times \frac{D_{\rm O}}{N_{\rm H}}}$ and ${\bf W}^{O}\in{\mathbb R}^{D_{\rm O}\times {D_{\rm O}}}$ are learnable weights; $N_{\rm H}$ is the number of heads. The ${\rm Attention}(\cdot)$ function is given by
	\begin{flalign}
		&{\rm Attention}({\bar {\bf Q}}, {\bar {\bf K}}, {\bar {\bf V}}) \!=\! {\rm Softmax}\!\left(\!{{\bar {\bf Q}}{\bar {\bf K}}^T}\!/\!{\sqrt{{D_{\rm I}}/{N_{\rm H}}}}\right)\!{\bar {\bf V}},&
		\label{PMA Attention}
	\end{flalign}
	where ${\rm Softmax}$ is performed at the second dimension.
	In summary, the process of ${\rm PMA}(\cdot)$ are given by \eqref{PMA eq1}-\eqref{PMA Attention}. 
	
	It is easy to prove that the parameterized invariant function ${\rm PMA}:\mathbb{R}^{M\times D_{\rm I}}\to \mathbb{R}^{1\times D_{\rm O}}$ satisfies the invariance. Similar to ${\rm FC}_{\rm HOE}(\cdot)$, ${\rm PMA}(\cdot)$ can be applied to a certain dimension of a tensor with batch computation. 
	The computational complexity of ${\rm PMA}(\cdot)$ mainly resides in \eqref{MAB eq1}-\eqref{PMA Attention}, denoted as ${\mathcal O}({\bar M}D_{\rm O}(D_{\rm I}+D_{\rm O}))$, with the number of parameters ${\mathcal O}(D_{\rm O}(D_{\rm I}+D_{\rm O}))$. 
	
	According to the definition of multidimensional invariance, we construct the module ${\rm MDI}_{\mathcal P}(\cdot)$, which applies distinct PMA across each dimension in the set ${\mathcal P}$. For example, to model an invariant mapping $f:\mathbb{R}^{M_1\times \cdots\times M_N\times D_X}\to \mathbb{R}^{D_Y}$, let ${\mathcal P}={\mathcal N}$, and its expression is given by
	\begin{align}
		{\rm MDI}_{{\mathcal P}}(\tX) = [{\rm PMA}_1\circ_{1}\cdots{\rm PMA}_N\circ_{N}\tX],
	\end{align}
	where ${\rm PMA}_n$ refers to the $n$-th PMA module, $\circ_n$ denotes the computation of PMA along the $n$-th dimension, and the brackets $[\cdot]$ indicate the removal of empty dimensions of size 1 generated during this process.
	
	\subsection{Advantages of TENN Modules}\label{Exploit PE sec}
	The TENN modules designed in Sections \ref{PE linear layer sec}-\ref{IE Attention layer sec} satisfy TE that aligns with mappings mentioned in Section \ref{PE optimization problem sec}. Considering conventional NNs can approximate almost any mapping \cite{hornik1989multilayer}, a natural question arises: \textit{Why should we exploit TE for NN design?} Based on the analysis in Sections \ref{PE linear layer sec}-\ref{IE Attention layer sec}, we provide several reasons as follows:
	\begin{itemize}
		\item \textbf{Parameter sharing:} 
		The TENN modules lead to specific parameter sharing patterns \cite{hartford2018deep, ravanbakhsh2017equivariance}, greatly reducing the number of parameter. Furthermore, the parameter count is independent of input size, which provides advantages for scenarios that involve a large number of items \cite{10198239}.
		\item \textbf{Lower complexity:} Under the parameter-sharing pattern, the restructuring of network operations (e.g., the multidimensional equivariant module discussed in Section \ref{PE linear layer sec}) reduces the computational complexity of inference. As the input size increases, the rate of complexity growth is relatively slow.
		\item \textbf{Flexible input size:} Since the parameters of equivariant networks are independent of the number of inputs, the network can work in scenarios with different input sizes without any modification \cite{zaheer2017deep, maron2018invariant}. 
		\item \textbf{Widespread presence:} It is easily demonstrated that the design of modulation \cite{mukhtar2012adaptive}, soft demodulation \cite{wang2023soft}, detection \cite{9298921}, channel estimation \cite{8481590} (or other parameter estimation), and other aspects also involve TE. Moreover, the dimensionality of these properties grows with the increases of device types in the system, such as access points, reconfigurable intelligent surfaces, and unmanned aerial vehicles. 
	\end{itemize}
	
	\section{Tensor Equivariance Framework \\for NN Design}\label{NN design sec}
	
	In this section, by leveraging the plug-and-play TENN modules, we first present the TE framework for NN design. Based on this framework, we construct NNs for solving optimization problems outlined in Sections \ref{precoding design sec} and \ref{PE HighOd layer sec}, respectively, as exemplified. 
	
	\subsection{Unified TE Framework}\label{method subsection}
	Firstly, we present the following proposition to establish the foundation for stacking equivariant layers, thus achieving multidimensional equivariance, high-order equivariance, and invariance in certain dimensions.
	\begin{ppn}\label{ppn stack}
		The high-order equivariant layers and multidimensional invariant layers retain their properties when stacked with multidimensional equivariant layers in front of them. 
	\end{ppn}
	\begin{pf}
		See Appendix \ref{pf stack}.
	\end{pf}
	The above proposition applies not only to the proposed modules but also to any other AI module that satisfies the properties. Building upon this proposition, we propose the following design framework.
	\begin{enumerate}
		\item \textbf{Find TE:} 
		Similar to Section \ref{precoding design sec} and Section \ref{US design sec}, formulate the optimization problem, construct the target mapping from the available information to the optimal solution, and then derive the TE that the target mapping possesses.
		\item \textbf{{Construct $\tX$ and $\tY$:}} 
		Given the properties to be satisfied in each dimension, available tensors are manipulated through operations such as repetition and concatenation to construct the input $\tX$ of the network. Similarly, the desired output $\tY$ is constructed for the required tensors.
		\item \textbf{{Build equivariant network:}} Based on the TE required by the mapping from $\tX$ to $\tY$, select modules from those proposed in Section \ref{PE optimization problem sec}, and then stack them to form the equivariant NN. 
		\item \textbf{{Design the output layer:}} The schemes in wireless communication system are usually constrained by various limitations, such as the transmit power. Therefore, it is necessary to design the output layer of the network to ensure that the outputs satisfy the constraints.	
	\end{enumerate}
	It is noteworthy that most of the existing techniques applicable to the design of AI-assisted communication schemes remain relevant within this framework. For instance, the technique of finding low-dimensional variables in \cite{shi2023robust}, the approach for non-convex optimization problems presented in \cite{sun2018learning}, and the residual connection for deep NNs in \cite{he2016identity}. 
	
	\subsection{TENN Design for Precoding}\label{precoding network design sec}
	Compared to the precoding tensor $\tW$, the auxiliary tensors $\tA^\star, \tU^\star$ in the optimal closed-form expression have smaller size, and incorporating such expression as the model-driven component can reduce the difficulty of NN training. Therefore, in this section, we consider designing NN to approximate the mapping from $\tH,\sigma^2$ to $\tA^\star, \tU^\star$.
	
	\subsubsection{Find TE}
	According to Section \ref{precoding design sec}, the mapping satisfys the equivariance in \eqref{ppn CF precoding e1}-\eqref{ppn CF precoding e3}.
	\subsubsection{Construct the Input and Output}
	We construct the input and output of the equivariant network as follows
	\begin{gather}
		\tX \!=\! [\real(\tH), \imaginary(\tH), \sigma^2 {\bf 1}]_{4}\!\in\!\mathbb{R}^{K\times N_{\rm R}\times N_{\rm T}\times D_X},\\
		\tY \!=\! [\real(\tA^\star), \!\imaginary(\tA^\star), \!\real(\tU^\star), \imaginary(\tU^\star)]_{4}\!\in\!\mathbb{R}^{K\times N_{\rm R}\times N_{\rm R}\times D_Y},
	\end{gather}
	where $D_X=3$ and $D_Y=4$. The mapping $G(\cdot)$ from $\tX$ to $\tY$ satisfies the following properties
	\begin{align}
		&G({\pi_K}\!\circ_{1}\!\tX)=\pi_{K}\circ_1\tY,\  \forall {\pi}_K\in {\mathbb{S}}_K,\label{XY CF precoding e1}\\
		&G({\pi_{N_{\rm R}}}\!\circ_{2}\tX)=\pi_{N_{\rm R}}\circ_{[2,3]}\tY,\  \forall {\pi}_{N_{\rm R}}\in {\mathbb{S}}_{N_{\rm R}},\label{XY CF precoding e2}\\
		&G({\pi_{N_{\rm T}}}\!\circ_{3}\tX)=\tY,\  \forall {\pi}_{N_{\rm T}}\in {\mathbb{S}}_{N_{\rm T}}. \label{XY CF precoding e3}
	\end{align}
	\subsubsection{Build Equivariant Network}
\begin{figure}
		\centering
		\includegraphics[width=0.31\textwidth]{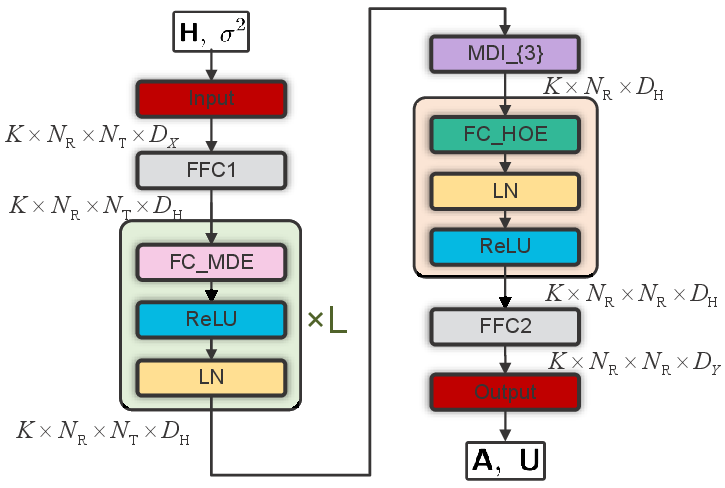}
		\caption{The architecture of TEPN.}
		\label{CFPN Architecture}
	\end{figure}

	The constructed network is illustrated in \figref{CFPN Architecture}. We first use the ${\rm FFC}_1(\cdot)$ to elevate the feature length of $\tX$ from $D_X$ to the hidden layer feature length $D_{\rm H}$. Since the desired function $G(\cdot)$ exhibits equivariance in the first three dimensions of $\tX$, we employ $L$ multidimensional ($N=3$) equivariant module ${\rm FC}_{\rm {MDE}}(\cdot)$ from Section \ref{PE linear layer sec} to perform the interaction between features of $\tX$. Furthermore, considering the invariance of $G(\cdot)$ in \eqref{XY CF precoding e3}, we employ the module ${\rm MDI}_{\{3\}}(\cdot)$ from Section \ref{IE Attention layer sec}, which satisfies the invariance, on the third dimension. To satisfy the high-order equivariance in \eqref{XY CF precoding e2}, we apply the high-order equivariant module ${\rm FC}_{\rm HOE}(\cdot)$ from Section \ref{PE HighOd layer sec} on the second dimension.
	$D_{\rm I}$ and $D_{\rm O}$ of all mentioned equivariant modules are equal to $D_{\rm H}$. Finally, we employ ${\rm FFC}_2(\cdot)$ to reduce the feature length from $D_{\rm H}$ to $D_{\rm Y}$.
	Between modules, we incorporate ReLU for element-wise nonlinearity, and adopt layer normalization (LN) to expedite training and improve performance \cite{ba2016layer}. 
	We refer to this network used for precoding as `TEPN'. 
	
	\subsubsection{Design the Output Layer} 
	$\tA$ and $\tU$ are key variables in the closed-form precoding expression, with constraints not explicitly shown. Therefore, the operations at the output layer are as follows
	\begin{align}
		\tA = \tY_{[:, :, :, 1]} + j\tY_{[:, :, :, 2]},\ \tU = \tY_{[:, :, :, 3]} + j\tY_{[:, :, :, 4]}.
	\end{align}
	By combining the decomposition of tensors into matrices and the concatenation of matrices into tensors, we can compute the final precoding scheme ${\hat \tW}$ from $\tA$ and $\tU$ using \eqref{precoding CF}.
	
	Given the optimization problem \eqref{3D precoding problem} for precoding, we employ unsupervised learning, with the negative loss function chosen as the objective function of problem \eqref{3D precoding problem}, i.e., 
	\begin{align}
		{\rm Loss} = -\frac{1}{N_{\rm sp}}\sum_{n=1}^{N_{\rm sp}}\sum_{k=1}^K R_k(\tH[n],{\hat \tW}[n],\sigma^2[n]),
	\end{align}
	where the subscript $[n]$ denotes the $n$-th sample in the dataset, and $N_{\rm sp}$ represents the number of samples.
	
	\subsection{TENN Design for User Scheduling}\label{US network design sec}
	\subsubsection{Find TE}
	According to Section \ref{precoding design sec}, the design of the user scheduling scheme targets to find the mapping from ${\tilde {\tH}},\sigma^2$ to ${\boldsymbol{\eta}}^\star$, which satisfys the equivariance in \eqref{ppn US e1}-\eqref{ppn US e3}.
	
	\subsubsection{Construct the Input and Output}
	We construct the input and output as follows
	\begin{gather}
		\tX = [\real({\tilde {\tH}}), \imaginary({\tilde {\tH}}), \sigma^2 {\bf 1}]_{4},\ 
		{\bf y} = {\boldsymbol{\eta}}^{\star}.
	\end{gather}
	The mapping $G(\cdot)$ from $\tX$ to ${\bf y}$ satisfies the following properties
	\begin{align}
		&G({\pi_{{\tilde K}}}\!\circ_{1}\!\tX)=\pi_{{\tilde K}}\circ_1{\bf y}, \  \forall {\pi}_{{\tilde K}}\in {\mathbb{S}}_{{\tilde K}},\label{XY US precoding e1}\\
		&G({\pi_{N_{\rm R}}}\!\circ_{2}\tX)={\bf y},\  \forall {\pi}_{N_{\rm R}}\in {\mathbb{S}}_{N_{\rm R}}, \label{XY US precoding e2}\\
		&G({\pi_{N_{\rm T}}}\!\circ_{3}\tX)={\bf y},\  \forall {\pi}_{N_{\rm T}}\in {\mathbb{S}}_{N_{\rm T}}. \label{XY US precoding e3}
	\end{align}
	
	\begin{figure}
		\centering
		\includegraphics[width=0.31\textwidth]{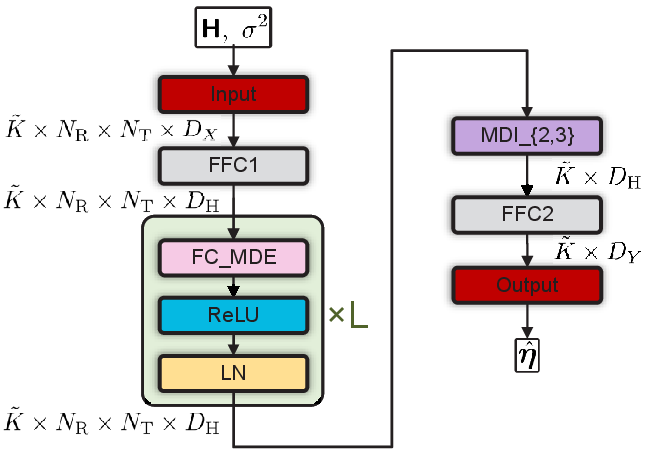}
		\caption{The architecture of TEUSN.}
		\label{USN Architecture}
	\end{figure}
	\subsubsection{Build Equivariant Network}
	The constructed network is illustrated in \figref{USN Architecture}. The overall structure is similar to TEPN. The difference lies in replacing ${\rm FC}_{\rm HOE}(\cdot)$ with another ${\rm MDI}_{\{2, 3\}}(\cdot)$ to satisfy invariance in the second and third dimensions. We refer to this network used for user scheduling as `TEUSN'. 
	
	\subsubsection{Design the Output Layer} 
	In problem \eqref{user scheduling problem}, ${\boldsymbol{\eta}}$ is constrained as a binary variable with all elements summing up to $K$. To address this, we first employ ${\rm Softmax}(\cdot)$ to transform the output $\tY$ into probabilities between 0 and 1, i.e., ${\boldsymbol{\eta}}^{\rm pro} = {\rm Softmax}({\bf y})$. Subsequently, the largest $K$ elements are set to 1, while the rest are set to 0, which is further denoted by ${\hat {\boldsymbol{\eta}}}$.
	
	We employ supervised learning, utilizing the following binary cross-entropy loss for training
	\begin{align}
		{\rm Loss}=\!-\!\frac{1}{N_{\rm sp}{\tilde K}}\sum_{n=1}^{N_{\rm sp}}\sum_{k=1}^{\tilde K} {\rm BCE}(\eta^{\rm pro}_{k}[n], \eta^{\star}_k[n]),
	\end{align}
	where ${\boldsymbol{\eta}}^{\star}$ is the target result, and ${\rm BCE}(a, b) = a\log(b)+(1-a)\log(1-b)$.

	\section{Numerical Results}\label{result sec}
	In this section, we employ the Monte Carlo method to assess the performance of the proposed methods. We consider a massive MIMO system and utilize QuaDRiGa channel simulator to generate all channel data \cite{6758357}. The configuration details of the channel model are as follows: The BS is equipped with uniform planar array (UPA) comprising $N_{\rm Tv}=2$ dual-polarized antennas in each column and $N_{\rm Th}$ dual-polarized antennas in each row with the number of antennas $N_{\rm T}=2N_{\rm Tv}N_{\rm Th}$. UEs are equipped with uniform linear array comprising $N_{\rm Rv}=1$ antennas in each column and $N_{\rm Rh}$ antennas in each row with the number of antennas $N_{\rm R}=N_{\rm Rv}N_{\rm Rh}$. In this section, parameters $N_{\rm Th}$ and $N_{\rm Rh}$ are adjusted to accommodate the desired antenna quantity configuration. Both the BS and UEs employ antenna type `3gpp-3d', the center frequency is set at 3.5 GHz, and the scenario is `3GPP\_38.901\_UMa\_NLOS' \cite{3GPPTR38.901}. Shadow fading and path loss are not considered. The cell radius is 500 meters, with users distributed within a 120-degree sector facing the UPA (3-sector cell). For the convenience of comparison, we consider the normalized channel satisfying $\sum_{k=1}^{K}{\rm Tr}\{{\bf H}_k{\bf H}^H_k\}=KN_{\rm R}N_{\rm T}$ and ${\rm SNR}=P_{\rm T}/{\sigma^2}$ \cite{5673745}. Under the same channel model configuration, all channel realizations are independently generated, implying diversity in the channel environments and terminal locations.
	
	\subsection{Training Details}
	For the network TEPN constructed for precoding in Section \ref{precoding network design sec}, its channel dataset size is $[60000, K, N_{\rm R}, N_{\rm T}, 2]$, with $55000$ channels used for training and $5000$ channels used for testing, and the channels are stored as real and imaginary parts. Similarly, for the network TEUSN constructed for user scheduling in Section \ref{US network design sec}, its channel dataset size is $[60000, {\tilde K}, N_{\rm R}, N_{\rm T}, 2]$. Besides the label (${\boldsymbol{\eta}}^\star$) dataset size is $[60000, {\tilde K}, 1]$, where ${\boldsymbol{\eta}}^\star$ is generated by well-performing conventional scheduling algorithms, as will be discussed in subsequent sections. We employ the same training strategy for TEPN and TEUSN. The number of iterations and batch size are set to be $2\times 10^{5}$ and 2000. We utilize the Adam optimizer with a learning rate of $5\times 10^{-4}$ for the first half of training and $5\times 10^{-5}$ for the latter half \cite{kingma2014adam}. 
	It should be noted that during training for TEPN, we provided data with noise under different SNR levels, randomly drawn from $\{0, 5, 10, 15, 20, 25, 30, 25, 40\}$. This enables the network to operate effectively across various SNRs, while achieving a good trade-off between performance and storage overhead.
	Similarly, data at SNRs $\{0, 10, 20, 30, 40\}$ are used for training TEUSN.
	Considering that the training of TEUSN requires labeled data, we can achieve a tradeoff between performance and training by reducing the diversity of SNR values and the number of channel samples in the dataset.
	
	\subsection{Performance of Precoding Schemes}\label{prec performance sec}
	
	This section compares the following methods:
	\begin{itemize}
		\item `\textbf{ZF}' and `\textbf{MMSE}': Conventional closed-form linear precoding methods \cite{1391204}.
		\item `\textbf{WMMSE-RandInt}' and `\textbf{WMMSE-MMSEInit}': Conventional algorithms for iterative solving of the sum-rate maximization precoding problem \cite{4712693}. `-RandInt' and `-MMSEInit' 
		represent using random tensor and MMSE precoding as initial values, respectively. The maximum number of iterations is set to 300 and the stopping criterion is defined as a reduction in the sum-rate per single iteration being less than $10^{-4}$. Similar to most precoding studies, we adopt WMMSE as our state-of-the-art baseline.
		\item `\textbf{GNN}': The AI-aided approach utilizing GNN for computation of precoding tensors from CSI \cite{liu2023multidimensional}, where the number of hidden layers is $4$ and the number of hidden layer neurons is $D_{G}=128$.
		\item `\textbf{DNN}': The model-driven AI-aided precoding method based on fully-connected deep NN \cite{kim2020deep}.
		\item `\textbf{CNN}': The AI-aided precoding method utilizes the CNN network in \cite{zhang2022deep} to obtain auxiliary variables in the optimal closed form of precoding. 
		\item `\textbf{TECFP}': The precoding scheme based TEPN in Section \ref{precoding network design sec} with $L=3$ and $D_{\rm H}=8$. 
	\end{itemize}
	
	
	\begin{table}[t]
		\centering
		\caption{Computational Complexity of Precoding Schemes}
		\resizebox{0.9\columnwidth}{!}{
			\begin{tabular}{ccc}
				\toprule
				Methods & Complexity Order & Multiplications\\
				\midrule
				ZF  &   ${\mathcal O}(N_{\rm T}K^2N_{\rm R}^2)$  & $6.1\times 10^4$ \\
				MMSE &  ${\mathcal O}(N_{\rm T}K^2N_{\rm R}^2)$   &  $6.1\times 10^4$\\
				WMMSE-RandInt &  ${\mathcal O}(T_{\rm P1}N_{\rm T}K^2N_{\rm R}^2)$   &  $3.0\times 10^7$ \\
				WMMSE-MMSEInit &  ${\mathcal O}(T_{\rm P2}N_{\rm T}K^2N_{\rm R}^2)$   &  $3.1\times 10^7$ \\
				GNN &  ${\mathcal O}(N_{\rm T}KN_{\rm R}D_{G}^2)$   &  $1.0\times 10^8$    \\
				DNN &  ${\mathcal O}(N_{\rm T}KN_{\rm R}D_{D}+D^2_{D})$   &  $8.5\times 10^5$    \\
				CNN &  ${\mathcal O}(N_{\rm T}KN_{\rm R}N^2_{C}+N_{\rm T}K^2N^3_{\rm R}N_{C})$   &  $2.9\times 10^6$    \\
				TECFP & ${\mathcal O}(N_{\rm T}KN_{\rm R}D_{\rm H}^2+N_{\rm T}K^2N_{\rm R}^2)$    &   $1.0\times 10^6$ \\
				\bottomrule
			\end{tabular}%
		}
		\label{Precoding Complexity Table}%
	\end{table}%
	
	\begin{figure}[tbp]
		\centering
		\subfigure[$N_{\rm T}=24$, $K=6$.]{
			\includegraphics[width=2.5in]{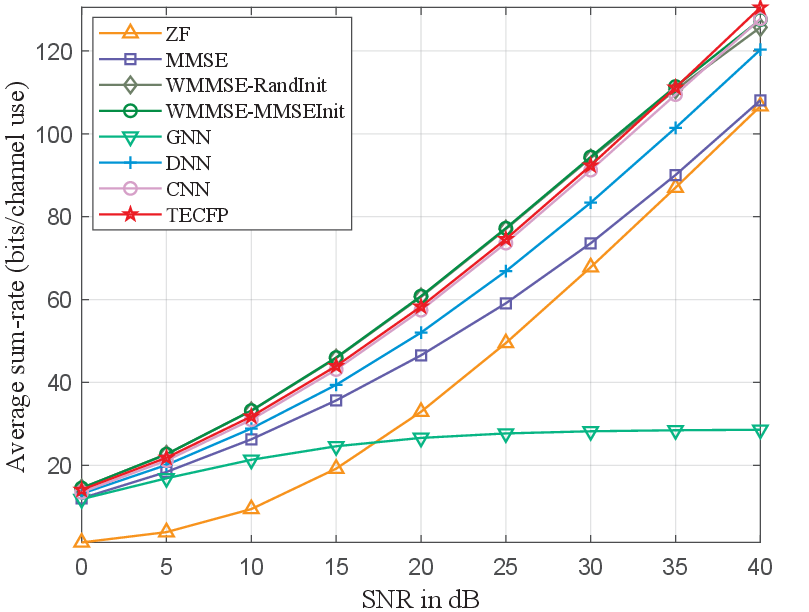}
			\label{sumrate 24mul6mul2}
		}
		\subfigure[$N_{\rm T}=32$, $K=8$.]{
			\includegraphics[width=2.5in]{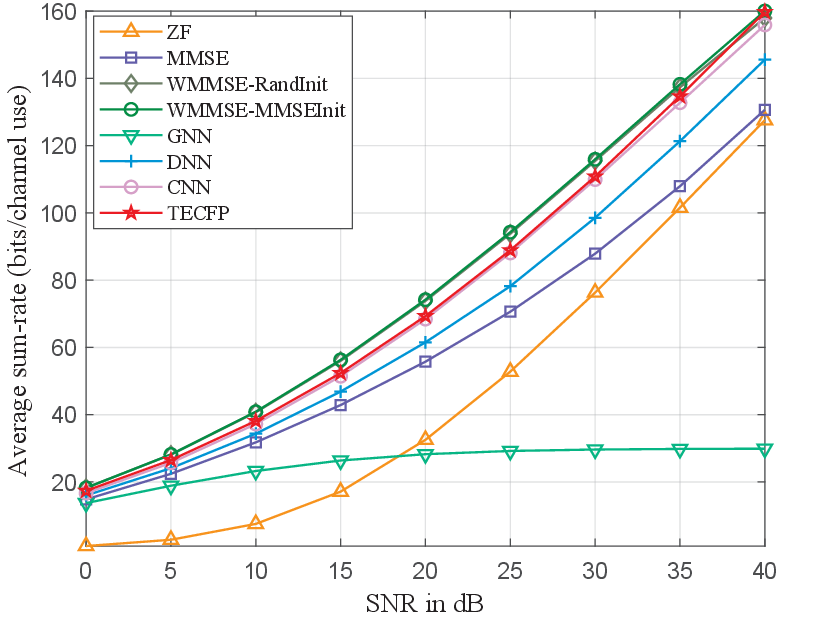}
			\label{sumrate 32mul8mul2}
		}
		\DeclareGraphicsExtensions.
		\caption{Sum rate vs SNR, $N_{\rm R}=2$.}
		\label{sumrate precoding fig}
	\end{figure}
	
	\begin{figure}[htbp]
		\centering
		\subfigure[Generalization for different $N_{\rm T}$.]{
			\includegraphics[width=2.5in]{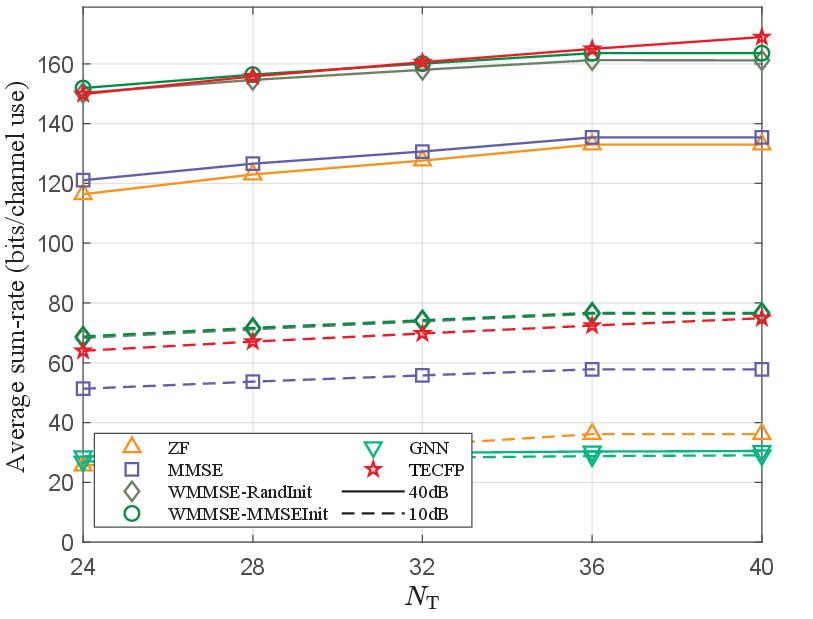}
			\label{CISB-NPA-MI}
		}
		\subfigure[Generalization for different $K$.]{
			\includegraphics[width=2.5in]{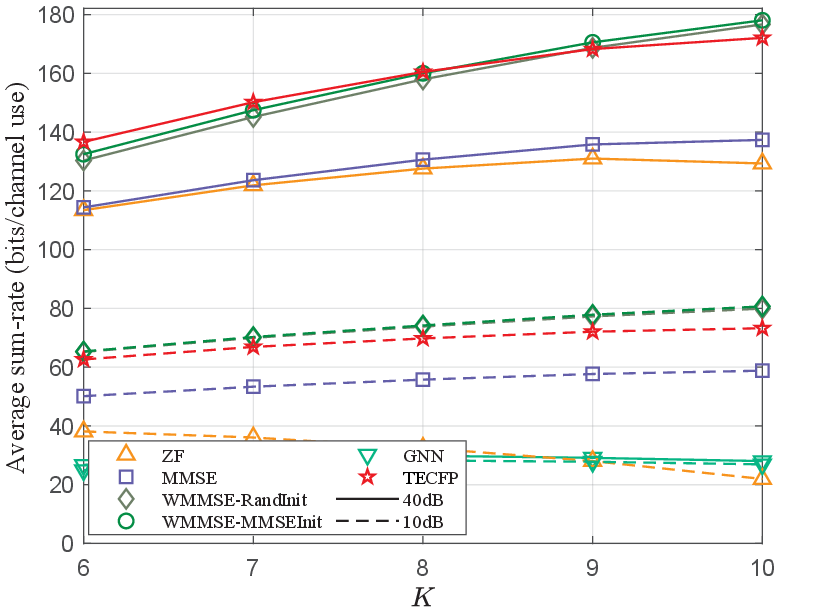}
			\label{CIMMSE-NPA-MI}
		}
		\DeclareGraphicsExtensions.
		\caption{Performance generalization of the network trained under scenario $N_{\rm T}=32$, $K=8$, and $N_{\rm R}=2$.}
		\label{generalization precoding}
	\end{figure}
	
	Table \ref{Precoding Complexity Table} contrasts the computational complexities of several methods under scenario $N_{\rm T}=32, K=8, N_{\rm R}=2$. ``Multiplications" refers to the count of real multiplications, with complex multiplications calculated as three times the real ones. It is noteworthy that in the table, $D_{D}>T_{\rm P1}\approx T_{\rm P2}>D_{\rm G}\gg N_{\rm T} > K \approx D_{\rm H} \approx N_{\rm C} > N_{\rm R}$, where $D_D$ denotes the number of neurons in the hidden layers of the DNN, and $N_{\rm C}$ denotes the number of convolutional kernels in the CNN. Among the considered methods, the complexity of ZF and MMSE precoding, as closed-form linear precoding methods, is the lowest. Although the computational complexity per single iteration of WMMSE precoding shares the same order as MMSE, achieving optimal performance typically requires multiple iterations, introducing high complexity. GNN also exhibit weight-sharing properties, thus their complexity is solely related to the first order of the channel dimensions. However, due to the high dimensionality of the precoding tensor, the approximation for precoding computations requires a substantial number of neurons $D_{G}$, thus introducing substantial complexity. TECFP leverages TE in mappings from CSI to auxiliary tensors, while enjoying the advantage of complexity being solely related to the first order of the channel size and requiring fewer neurons, thereby significantly reducing complexity. The required number of multiplications also validate the aforementioned analysis. It should be noted that, considering that network training can be performed offline, both the computational complexity presented in tabular form and the subsequent FLOPs comparisons only consider inference stage. In addition, the iterative process of non-deep learning methods is explicitly incorporated, contributing to their inherently higher computational complexity compared to AI-assisted approaches.
	Beyond hardware resources, the computational efficiency of different methods also depends on the optimization of their underlying implementations, which constitutes a key research direction within the TE framework in the future.
	
	

	The comparison of sum-rate for each precoding scheme in scenarios $N_{\rm T}=32, K=8, N_{\rm R}=2$ and $N_{\rm T}=24, K=6, N_{\rm R}=2$ is illustrated in \figref{sumrate precoding fig}. It can be observed that the overall sum-rate performance of precoding schemes with lower computational complexity, such as ZF and MMSE, is significantly lower than that of WMMSE, validating the superior performance of WMMSE. Despite GNN employing a large number of parameters and computational complexity, their performance remains poor. This is attributed to the necessity of matrix inversion for high-dimensional matrices during the computation of precoding matrices \cite{zhao2022learning}. In contrast, our proposed method integrates the advantages of model-driven approaches within the TE framework, enabling the approximation of WMMSE performance while maintaining low complexity. The performance of TECFP occasionally surpasses WMMSE at high SNR levels, possibly due to the heuristic nature of the WMMSE algorithm, which may result in suboptimal performance under high SNR conditions. \figref{generalization precoding} demonstrates the generalization capability of the proposed approach. We train our network in scenario $N_{\rm T}=32, K=8, N_{\rm R}=2$ and directly apply it to various different scenarios. It can be observed that the proposed method exhibits consistently outstanding performance, highlighting its robust practical utility.

	\begin{figure*}[t]
		\centering
		\begin{minipage}[t]{0.32\textwidth}
			\centering
			\includegraphics[width=1\textwidth]{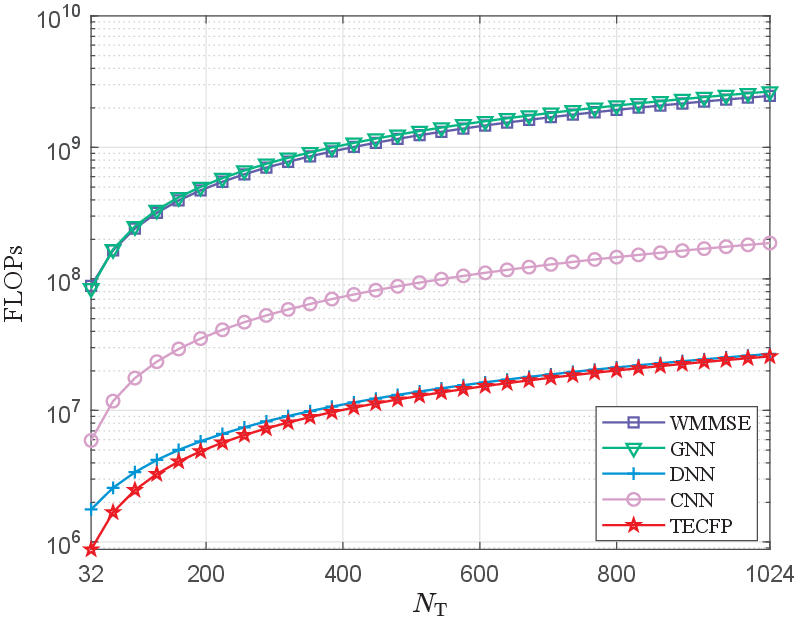}
			\caption{FLOPs vs $N_{\rm T}$, $K=8$, $N_{\rm R}=2$.}
			\label{FLOPs_txAntNum}
		\end{minipage}
		\begin{minipage}[t]{0.32\textwidth}
			\centering
			\includegraphics[width=1\textwidth]{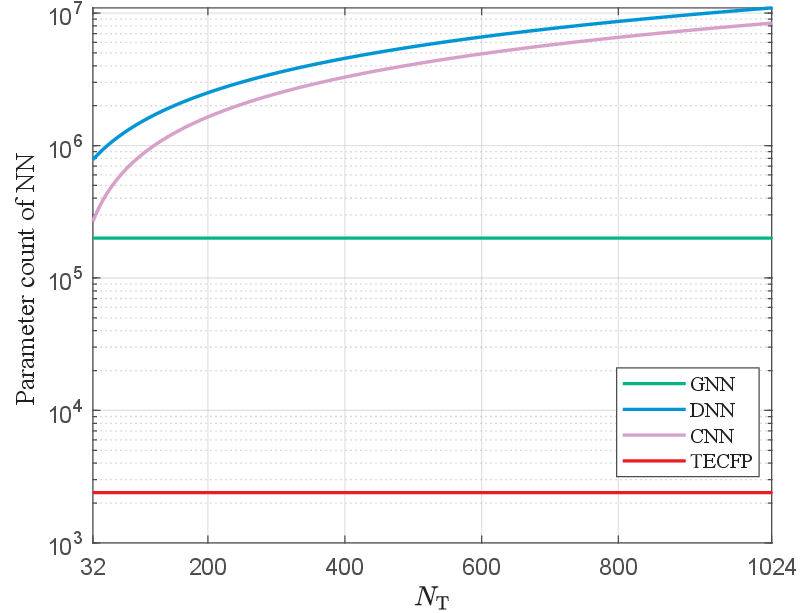}
			\caption{Parameter counts vs $N_{\rm T}$, $K=8$, $N_{\rm R}=2$.}
			\label{parameter_txAntNum}
		\end{minipage}
		\begin{minipage}[t]{0.32\textwidth}
			\centering
			\includegraphics[width=1\textwidth]{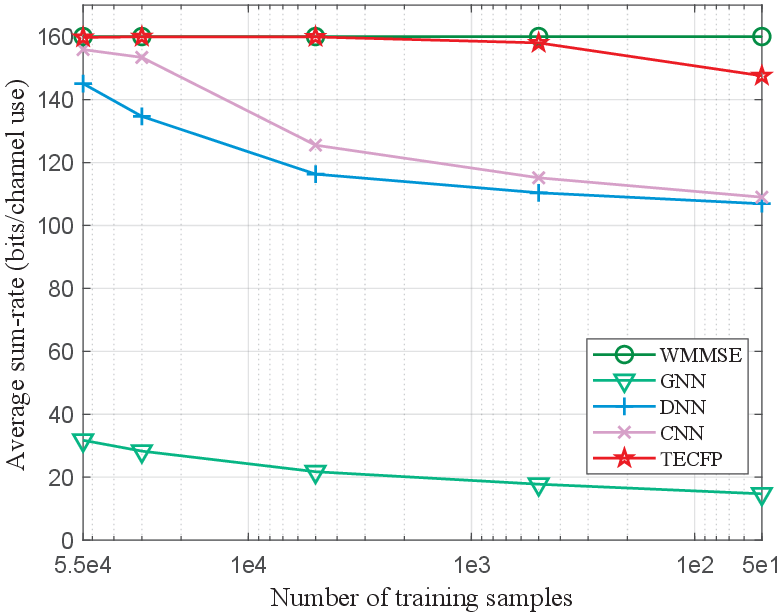}
			\caption{Performance vs training set size, $N_{\rm T}=32$, $K=8$, $N_{\rm R}=2$, $\text{SNR}=40\text{dB}$.}
			\label{dataset_40dB}
		\end{minipage}
	\end{figure*}

	We compared the FLOPs with different $N_{\rm T}$ in \figref{FLOPs_txAntNum}. For different $N_{\rm T}$, we only made necessary adjustments to the network architecture. Our method demonstrates lower FLOPs across various configurations, while its FLOP growth rate approaches that of DNN. The reason is that we did not adjust the number of neurons in the hidden layers of DNN. However, the performance of DNN is significantly influenced by the number of neurons in the hidden layers.
	\figref{parameter_txAntNum} compares the changes in parameter count required by the NN to match input and output dimensions when the number of transmit antennas is adjusted. As the number of antennas increases, the FC layers in CNNs and DNNs require corresponding adjustments, resulting in an increase in parameter count. In contrast, both our network and GNNs do not require any adjustments in parameter count, highlighting the advantages of the proposed method in handling large-scale problems. 
	\figref{dataset_40dB} compares the performance of the network when the training set contains different numbers of samples ($\tH$). Without loss of generality, we maintain a consistent number of mini-batches across training on different dataset sizes. It can be observed that the performance of the proposed method declines at the slowest rate as dataset size decreases. This suggests that the proposed method requires a smaller dataset and has a stronger resistance to overfitting. This advantage stems from leveraging TE under the model-driven approach.
	
	\begin{figure}[htbp]
		\centering
		\includegraphics[width=3.2in]{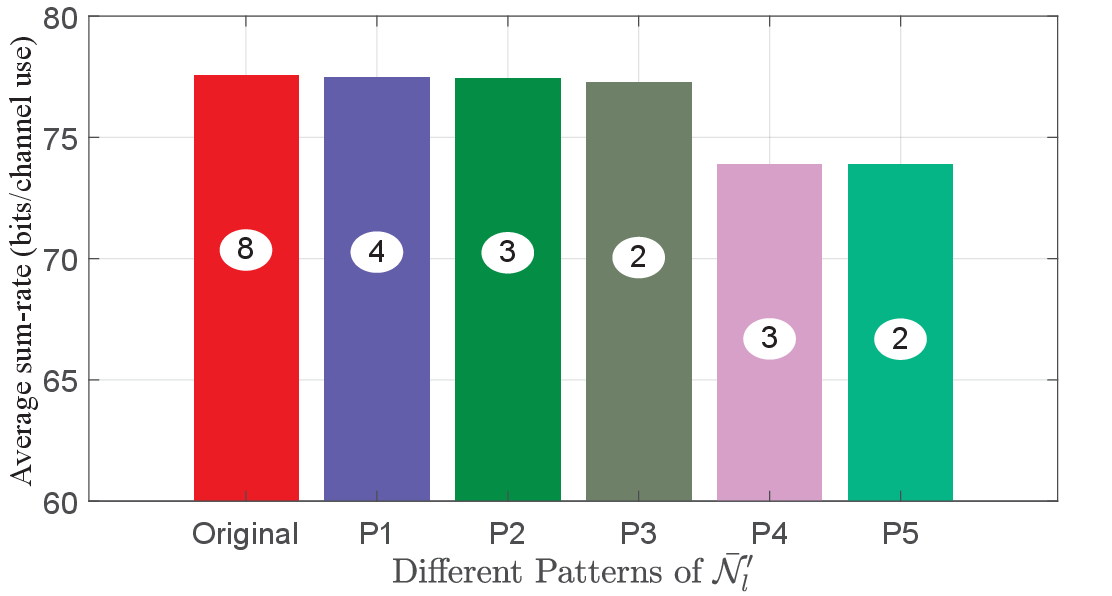}
		\caption{Comparison of performance and complexity for multidimensional equivariance modules with different low-complexity patterns (set ${\bar {\mathcal N}'}_l$). The numbers on the bars represent the relative complexity.}
		\label{low-complexity pattern}
	\end{figure}
	
	Another interesting topic is the method proposed in Section \ref{PE linear layer sec} for further reducing the complexity of the multidimensional and high-order equivariant modules. For the multidimensional equivariance module, the main idea is constructing a subset ${\bar {\mathcal N}}'_l$ to replace ${\bar {\mathcal N}}$ for the $l$-th layer of the network. In Section \ref{PE linear layer sec}, we provide three patterns with examples. The multidimensional equivariance module in the TECFP method is reconstructed according to these three examples, denoted as P1, P2, and P3. For comparison, we also present two additional patterns, P4: ${\bar {\mathcal N}}^{\rm P4}_{l} = \{\varnothing, \{1\},\{3\}\}, \forall l$ and P5: ${\bar {\mathcal N}}^{\rm P5}_{l} = \{\varnothing, \{1\}\}, \forall l$. In \figref{low-complexity pattern} along with TECFP (`Original'), we compare the sum rate of these methods in the scenario $N_{\rm T}=32$, $K=8$, $N_{\rm R}=2$, which is averaged over SNRs of $\{0,5,...,40\}$. In \figref{low-complexity pattern}, the numbers on the bars represent the relative computational complexity and parameter counts of the modules under several patterns. For instance, under pattern P3, the computational complexity and parameter count of this module are $25\%$ of those (`Original') in TECFP. It can be observed that the proposed patterns significantly reduce the complexity while ensuring good performance. In contrast, P4 and P5 do not satisfy \eqref{pattern condition}, which results in a loss of the ability to capture global features across all dimensions, leading to a significant decline in performance.

	\begin{table}[t]
		\centering
		\caption{Computational Complexity of User Scheduing Schemes}
		\resizebox{1.0\columnwidth}{!}
		{
			\begin{tabular}{ccc}
				\toprule
				Methods & Complexity Order & Multiplications \\
				\midrule
				MMSE-Rand &  ${\mathcal O}(N_{\rm T}K^2N_{\rm R}^2)$   &   $6.1\times 10^4$\\
				MMSE-Greedy  &   ${\mathcal O}(K{\tilde K}(N_{\rm T}K^2N_{\rm R}^2))$  &  $1.8\times 10^{6}$ \\
				MMSE-TEUS &  ${\mathcal O}(N_{\rm T}{\tilde K}N_{\rm R}D_{{\rm H}1}^2+N_{\rm T}K^2N_{\rm R}^2)$   &   $1.4\times 10^{6}$  \\
				WMMSE-Rand &  ${\mathcal O}(T_{P2}N_{\rm T}K^2N_{\rm R}^2)$   &  $3.1\times 10^7$ \\
				WMMSE-Greedy &  ${\mathcal O}(K{\tilde K}(T_{P2}N_{\rm T}K^2N_{\rm R}^2))$   &  $6.4\times 10^{8}$  \\
				WMMSE-TEUS & ${\mathcal O}(N_{\rm T}{\tilde K}N_{\rm R}D_{{\rm H}2}^2+T_{P2}N_{\rm T}K^2N_{\rm R}^2)$    &  $5.8\times 10^{7}$ \\
				TECFP-TEUS & ${\mathcal O}(N_{\rm T}{\tilde K}N_{\rm R}D_{{\rm H}2}^2+N_{\rm T}K^2N_{\rm R}^2)$    &  $2.8\times 10^{7}$   \\
				\bottomrule
			\end{tabular}
		}
		\label{US Complexity Table}%
	\end{table}

	\subsection{Performance of User Scheduling Schemes}
	
	\begin{figure}[tbp]
		\centering
		\includegraphics[width=2.5in]{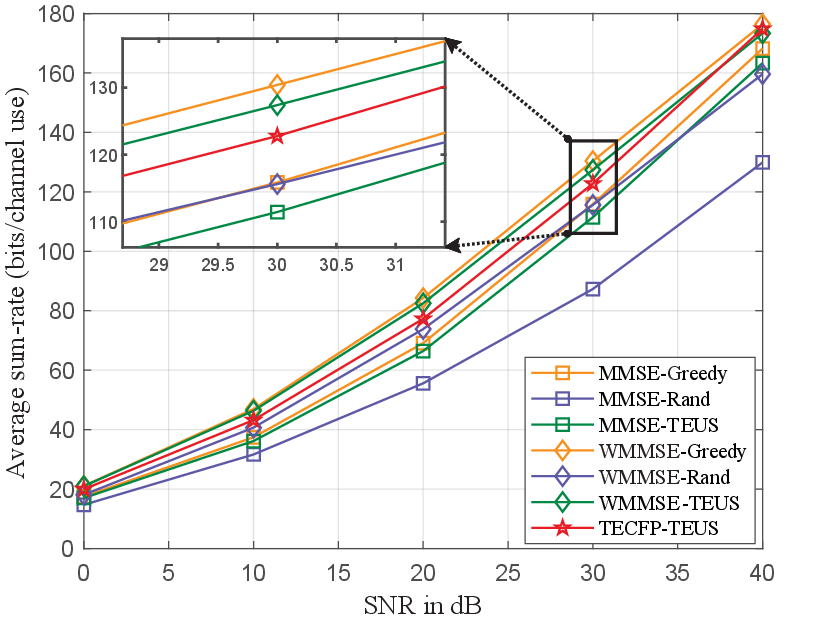}
		\caption{Sum rate vs SNR, $N_{\rm T}=32$, ${\tilde K}=12$, $K=8$, $N_{\rm R}=2$.}
		\label{US sumrate fig}
	\end{figure}
	\begin{figure}[htbp]
		\centering
		\subfigure[Generalization for different $N_{\rm T}$.]{
			\includegraphics[width=2.4in]{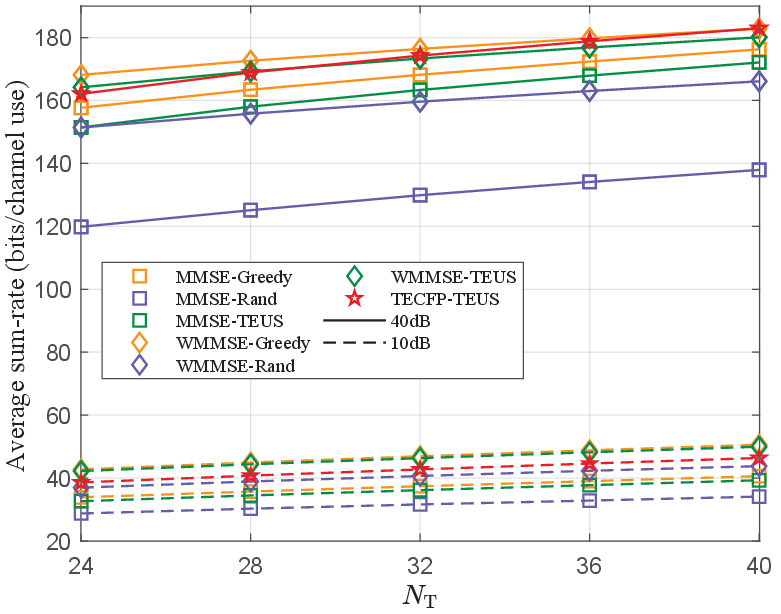}
			\label{US_GN_txAntNum_sumrate_fig}
		}
		\subfigure[Generalization for different ${\tilde K}$.]{
			\includegraphics[width=2.4in]{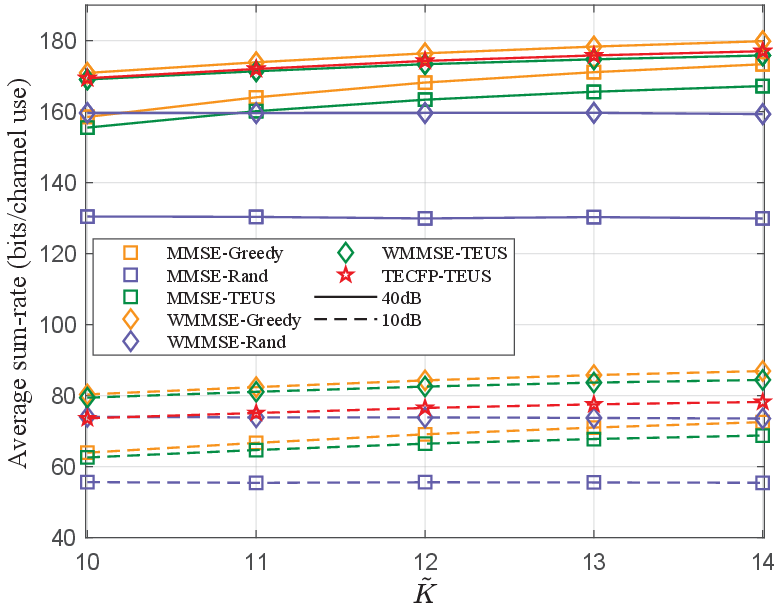}
			\label{US_GN_kNum_sumrate_fig-NPA-MI}
		}
		\DeclareGraphicsExtensions.
		\caption{Performance generalization of the network trained under scenario $N_{\rm T}=32$, ${\tilde K}=12$, $K=8$, and $N_{\rm R}=2$.}
		\label{generalization US}
	\end{figure}
	
	Based on precoding schemes MMSE, WMMSE (MMSEInit), and TECFP as a foundation, we compare several user scheduling strategies as follows:
	\begin{itemize}
		\item `\textbf{Rand}': Select $K$ users randomly from ${\tilde K}$ users.
		\item `\textbf{Greedy}': Select users one by one from ${\tilde K}$ users based on the criterion of maximizing the sum-rate after precoding for the selected users until reaching $K$ users \cite{dimic2005downlink}.
		\item `\textbf{TEUS}': The scheduling strategy based on TEUSN trained with the result of greedy scheduling strategy as the label in Section \ref{US network design sec}.
	\end{itemize}
	The scheduling strategies vary among different precoding schemes. It is worth noting that the results of greedy user scheduling vary across different precoding schemes. The number of hidden layers and nodes in the TEUS used for MMSE and WMMSE are respectively denoted as $L=3$, $D_{{\rm H}}=8$ and $L=4$, $D_{{\rm H}}=32$.
	
	We compare the computational complexity of several scheduling and precoding combination schemes in Table \ref{US Complexity Table} under scenario $N_{\rm T}=32$, ${\tilde K}=12$, $K=8$, $N_{\rm R}=2$. It can be observed that although the greedy scheduling algorithm is designed to select the near-optimal users, it introduces a high computational complexity. The proposed method, TEUS, significantly reduces the overall computational complexity of scheduling and precoding.  \figref{US sumrate fig} compares the sum-rate performance of various precoding and scheduling combination schemes under scenario $N_{\rm T}=32$, ${\tilde K}=12$, $K=8$, $N_{\rm R}=2$. It can be observed that the performance of MMSE-TEUS and WMMSE-TEUS is close to that of MMSE-Greedy and WMMSE-Greedy, respectively. This indicates that the proposed TEUS can achieve outstanding performance at lower computational complexity. Although the performance of TECFP-TEUS is slightly inferior to that of the proposed WMMSE-TEUS, its computational complexity is much lower, requiring less than 5\% of that of WMMSE-Greedy. Furthermore, in \figref{generalization US}, we compare the generalization ability of the proposed method across different scenarios. It is evident that, as the scenario changes, the performance trends of MMSE-TEUS, WMMSE-TEUS, and TECFP-TEUS are similar to those of the conventional algorithms MMSE-Greedy and WMMSE-Greedy. This implies that these proposed schemes possess outstanding capability to be directly extended for application in different scenarios.

	\section{Conclusion}\label{conclusion}
	In this paper, we proposed the unified TE framework, leveraging TE in MU-MIMO systems. Firstly, we defined the concept of TE, which encompasses definitions of multiple properties. On this basis, we put forward the TE framework, which is capable of designing NNs with TE for MU-MIMO systems. In this framework, the various modules are plug-and-play, allowing them to be stacked to accommodate different properties and applicable for various wireless communication tasks. Taking precoding and user scheduling problems as examples, we efficiently designed corresponding AI-assisted schemes using the proposed TE framework, which demonstrates that this framework can reduce the complexity of method design. The corresponding simulation results validates the superiority of the proposed modules and the unified TE framework. In future work, we will continue to optimize the underlying code of the proposed TE modules to improve their running efficiency.
	
	\appendices 
	\section{Proof of Proposition \ref{ppn precoding}}\label{pf precoding}
	We demonstrate partial conclusions in the proposition as follows 
	\begin{align}
		\sum_{k=1}^KR_k(\tH,\tW,\sigma^2) 
		&= \sum_{k=1}^KR_k(\pi_{K}\circ_1\tH,\pi_{K}\circ_1\tW,\sigma^2),\label{ppn1 prove e1}\\ 
		&= \sum_{k=1}^KR_k({\pi_{N_{\rm R}}}\!\circ_{2}\!\tH,{\pi_{N_{\rm R}}}\!\circ_{2}\!\tW,\sigma^2),\label{ppn1 prove e2}\\ &= \sum_{k=1}^KR_k({\pi_{N_{\rm T}}}\!\circ_{3}\!\tH,{\pi_{N_{\rm T}}}\!\circ_{3}\!\tW,\sigma^2).\label{ppn1 prove e3}
	\end{align}
	Then, the remaining conclusions in the proposition can be obtained through proof by contradiction.

	We first consider the proof of \eqref{ppn1 prove e1}. For convenience, we use $\pi$ to denote $\pi_{K}$, and the sum-rate expression considering its influence is as follows
	\begin{align}
	\begin{split}
	&R_k(\pi\circ_1\tH,\pi\circ_1\tW,\sigma^2)\\
	&=\log{\rm det}\left({\bf I}+{\bf W}_{\pi(k)}{\bf H}^H_{\pi(k)}({\boldsymbol {\Omega}}'_{k})^{-1}{\bf H}_{\pi(k)}{\bf W}^H_{\pi(k)}\right),
	\end{split}
	\end{align}
	where ${\boldsymbol {\Omega}}'_{k} = \sigma^2{\bf I}+\sum_{i=1,i\neq {\pi(k)}}^{K}{\bf H}_{\pi(k)}{\bf W}^H_i{\bf W}_i{\bf H}^H_{\pi(k)} = {\boldsymbol {\Omega}}_{\pi(k)}$.
	Thus, we have 
	\begin{align}
	&R_k(\pi\circ_1\tH,\pi\circ_1\tW,\sigma^2)\nonumber\\
	&=\log{\rm det}\left({\bf I}+{\bf W}_{\pi(k)}{\bf H}^H_{\pi(k)}{\boldsymbol {\Omega}}^{-1}_{\pi(k)}{\bf H}_{\pi(k)}{\bf W}^H_{\pi(k)}\right)\\
	&=R_{\pi(k)}(\tH,\tW,\sigma^2).\nonumber
	\end{align}
	Substituting this expression into \eqref{ppn1 prove e1} yields its validity.
	
	Subsequently, we consider the proof of \eqref{ppn1 prove e2} and use $\pi$ to denote $\pi_{N_{\rm R}}$. We define the permutation matrix ${\boldsymbol{\Pi}}$ to represent the permutation of $\pi$ at the second dimension of $\tH$ and $\tW$. In matrix ${\boldsymbol{\Pi}}$, each row contains only one element equal to 1, with all other elements being 0, and all elements 1 are located in distinct columns. Note that ${\boldsymbol{\Pi}}{\boldsymbol{\Pi}}^T={\bf I}$. 
	For $\pi\circ_{2}\tH$ and $\pi\circ_{2}\tW$, the corresponding channel and precoder of the $k$-th user can be expressed as ${\bf H}'_k={\boldsymbol{\Pi}}{\bf H}_k$ and ${\bf W}'_k={\boldsymbol{\Pi}}{\bf W}_k$. On this basis, we have
	\begin{align}
	\begin{split}
	&R_k(\pi\circ_2\tH,\pi\circ_2\tW,\sigma^2)\\
	&=\log{\rm det}\left({\bf I}+{\bf W}'_{k}({\bf H}'_{k})^H({\boldsymbol {\Omega}}'_{k})^{-1}{\bf H}'_{k}({\bf W}'_{k})^H\right)\\
	&=\log{\rm det}\left({\bf I}+{\boldsymbol{\Pi}}{\bf W}_{k}{\bf H}^H_{k}{\boldsymbol{\Pi}}^T({\boldsymbol {\Omega}}'_{k})^{-1}{\boldsymbol{\Pi}}{\bf H}_{k}{\bf W}_{k}^H{\boldsymbol{\Pi}}^H\right),
	\end{split}
	\end{align}
	where ${\boldsymbol {\Omega}}'_{k} = \sigma^2{\bf I}+\sum_{i=1,i\neq k}^{K}{\boldsymbol{\Pi}}{\bf H}_k{\bf W}^H_i{\boldsymbol{\Pi}}^T{\boldsymbol{\Pi}}{\bf W}_i{\bf H}^H_k{\boldsymbol{\Pi}}^T$. According to Sylvester determinant identity that  $\det({\bf I}+{\bf A}{\bf B})=\det({\bf I}+{\bf B}{\bf A})$, it can be derived that
	\begin{align} 
	\begin{split}
	&R_k(\pi\circ_2\tH,\pi\circ_2\tW,\sigma^2)\\
	&=\log{\rm det}\left({\bf I}+{\bf W}_{k}{\bf H}^H_{k}{\boldsymbol{\Pi}}^T({\boldsymbol {\Omega}}'_{k})^{-1}{\boldsymbol{\Pi}}{\bf H}_{k}{\bf W}_{k}^H\right).
	\label{ppn precoding prove sumrate1}
	\end{split}
	\end{align}
	According to Woodbury matrix identity, we have $({\boldsymbol {\Omega}}'_{k})^{-1}={\boldsymbol{\Pi}}({\boldsymbol {\Omega}}_{k})^{-1}{\boldsymbol{\Pi}}^T$.
	Substituting this expression into \eqref{ppn precoding prove sumrate1} yields $R_k(\pi\circ_2\tH,\pi\circ_2\tW,\sigma^2) = R_k(\tH,\tW,\sigma^2)$.
	Therefore, \eqref{ppn1 prove e2} holds. Similarly, it can be proved that \eqref{ppn1 prove e3} also holds.
	
	Based on \eqref{ppn1 prove e1}-\eqref{ppn1 prove e3}, Given a fixed $\sigma$, it is easy to prove that if $\tW^{\star}$ is one of the optimal solutions for problem \eqref{3D precoding problem} based on $\tH$, then $\pi_{K}\circ_1\tW^{\star}$, $\pi_{N_{\rm R}}\circ_2\tW^{\star}$, and $\pi_{N_{\rm T}}\circ_3\tW^{\star}$ are also ones of the optimal solutions for problem \eqref{3D precoding problem} based on $\pi_{K}\circ_1\tH$, $\pi_{N_{\rm R}}\circ_2\tH$, and $\pi_{N_{\rm T}}\circ_3\tH$, respectively. This completes the proof.
	
	\section{Proof of Proposition \ref{ppn CF precoding}}\label{pf CF precoding}
	Based on \ppnref{ppn precoding}, the validity of the following equation leads to the establishment of \ppnref{ppn CF precoding}. 
	\begin{flalign}
		&\pi_K\circ_1\tW = {\rm CFP}\left(\pi_K\!\circ_1\!\tH,\pi_K\!\circ_1\!\tA,\pi_K\!\circ_1\!\tU, \sigma^2\right),\label{ppn2 prove e1}\\
		&{\pi_{N_{\rm R}}}\!\circ_{2}\!\tW \!=\! {\rm CFP}\left({\pi_{N_{\rm R}}}\!\circ_{2}\!\tH,{\pi_{N_{\rm R}}}\!\circ_{[2, 3]}\!\tA,{\pi_{N_{\rm R}}}\!\circ_{[2, 3]}\!\tU, \sigma^2\right),\label{ppn2 prove e2}\\
		&{\pi_{N_{\rm T}}}\!\circ_{3}\!\tW \!=\! {\rm CFP}\left({\pi_{N_{\rm T}}}\!\circ_{3}\!\tH,\tA,\tU, \sigma^2\right),\label{ppn2 prove e3}
	\end{flalign}
	where $\tW={\rm CFP}\left(\tH,\tA,\tU\right)$. Next, we will separately prove these equations.

	We first consider the proof of \eqref{ppn2 prove e1}. We use $\pi$ to denote $\pi_{K}$ and define $\tW'={\rm CFP}\left(\pi_K\!\circ_1\!\tH,\pi_K\!\circ_1\!\tA,\pi_K\!\circ_1\!\tU\right)$. According to \eqref{precoding CF} and Woodbury matrix identity, we have ${\tilde{\bf W}}'_k \!=\! {\bf U}_{\pi(k)}^T {\bf A}^{*}_{\pi(k)}{\bf H}^{*}_{\pi(k)}\left( {\boldsymbol{\Upsilon}}+\mu'{\bf I}_{N}\right)^{-1}$, where ${\boldsymbol{\Upsilon}}=\sum_{k=1}^K{\bf H}^H_k{\bf A}^H_k{\bf U}_k{\bf A}_k{\bf H}_k$, $\mu'={{\rm Tr}\left({\bf U}'{\bf A}'{\bf A}'^H\right)}={{\rm Tr}\left({\bf U}{\bf A}{\bf A}^H\right)}=\mu$, ${\bf U}' = {\rm blkdiag}\{{\bf U}_{\pi(1)},...,{\bf U}_{\pi(K)}\}$, and ${\bf A}' = {\rm blkdiag}\{{\bf A}_{\pi(1)},...,{\bf A}_{\pi(K)}\}$.
	Then, we can concludes ${\tilde{\bf W}}'_k={\tilde{\bf W}}_{\pi(k)}$, which leads to $\tW' = \pi_K\circ_1 \tW$.
	
	Subsequently, we consider the proof of \eqref{ppn2 prove e2}. We use $\pi$ to denote $\pi_{N_{\rm R}}$. We define the permutation matrix ${\boldsymbol{\Pi}}$ to represent the permutation of $\pi$ at $\tH$, $\tA$, and $\tU$. 
	For $\pi\circ_{2}\tH$, $\pi\circ_{[2, 3]}\tA$, and $\pi\circ_{[2, 3]}\tU$. The corresponding channel and auxiliary tensors of the $k$-th user can be expressed as ${\bf H}'_k={\boldsymbol{\Pi}}{\bf H}_k$, ${\bf A}'_k={\boldsymbol{\Pi}}{\bf A}_k{\boldsymbol{\Pi}}^T$, and ${\bf U}'_k={\boldsymbol{\Pi}}{\bf U}_k{\boldsymbol{\Pi}}^T$. After permutation, the expression for the precoding matrix before scaling for the $k$-th user is given by ${\tilde{\bf W}}'_k 
	= {\boldsymbol{\Pi}}{\bf U}_k^T {\bf A}^{*}_k{\bf H}^{*}_k\left({\boldsymbol{\Upsilon}} +\mu'{\bf I}_{N}\right)^{-1}$, which leads to $\tW' = {\boldsymbol{\Pi}}{\tilde{\bf W}}_k=\pi\circ_2 \tW$. Similar to the proof of \eqref{ppn2 prove e1}, \eqref{ppn2 prove e3} can be proved. This completes the proof.

	\section{Proof of Proposition \ref{ppn US}}\label{pf US}
	Similar to Appendix \ref{pf precoding}, the establishment of \ppnref{ppn US} can be proven by demonstrating the validity of the following equations 
	\begin{gather}
		R_{\rm US}({\tilde \tH},{\boldsymbol{\eta}},\sigma^2) = R_{\rm US}({\pi_{\tilde K}}\!\circ_{1}\!{\tilde \tH},{\pi_{\tilde K}}\!\circ_{1}\!{\boldsymbol{\eta}},\sigma^2),\label{ppn3 prove e1}\\
		R_{\rm US}({\tilde \tH},{\boldsymbol{\eta}},\sigma^2) = R_{\rm US}({\pi_{N_{\rm R}}}\!\circ_{2}\!{\tilde \tH},{\boldsymbol{\eta}},\sigma^2),\label{ppn3 prove e2}\\
		R_{\rm US}({\tilde \tH},{\boldsymbol{\eta}},\sigma^2) = R_{\rm US}({\pi_{N_{\rm T}}}\!\circ_{3}\!{\tilde \tH},{\boldsymbol{\eta}},\sigma^2).\label{ppn3 prove e3}
	\end{gather}
	
	For \eqref{ppn3 prove e1}, we have
	\begin{flalign}
	&R_{\rm US}({\pi}\!\circ_{1}\!{\tilde \tH},{\pi}\!\circ_{1}\!{\boldsymbol{\eta}},\sigma^2)\!=\!\sum_{k'\in {\mathcal{K}}'}\!\!\!R_{k'}(\tH',G_{\rm CP}(\tH', \sigma^2),\sigma^2),&
	\label{ppn3 prove sumrate1}
	\end{flalign}
	where ${\mathcal{K}}'\!=\!\{\pi(k)|\eta_{k}\!=\!1,{k}\!\in\!{{\tilde{\mathcal{K}}}}\}$ and $\tH' = [{\bf H}'_{k'}]_{0,k'\in{\mathcal{K}}'}\in\mathbb{C}^{K\times N_{\rm R}\times N_{\rm T}}$.
	It is easy to verify that ${\bf H}'_{k'}={\bf H}_{\pi^{-1}(k')}$. With the definition of ${\mathcal{K}}'$, we can conclude that $\tH'=\pi_K\circ_1\tH,\pi_K\in\mathbb{S}_K$. Substituting this equation into \eqref{ppn3 prove sumrate1} yields the establishment of \eqref{ppn3 prove e1}. Similarly, it can be proved that \eqref{ppn3 prove e2} and \eqref{ppn3 prove e3} also holds. This completes the proof.
	
	\section{Proof of Proposition \ref{ppn linear PE}}\label{pf linear PE}
	We prove the validity of this proposition under the scenario of $D_{\rm I}=D_{\rm O}=1$, and this conclusion can be easily extended to the scenario of $D_{\rm I}>1$ and $D_{\rm O}>1$. Besides, we temporarily ignore the bias ${\bf b}$ and derive its pattern at the end. We reshape the weights to $\tW\in\mathbb{R}^{M_1\times\cdots\times M_N\times M_1\times\cdots\times M_N}$ and use $\tW_{[{\bf p},{\bf q}]}$ to represent $\tW_{[p_1,...,p_N,q_1,...,q_N]}$. According to \cite{hartford2018deep}, it can be derived that the weights satisfying multidimensional equivariance across dimensions in ${\mathcal{N}}=\{1, ..., N\}$ exhibit the following pattern
	\begin{flalign}
	&\tW_{[{\bf p}, {\bf q}]}\!=\! w_{{\mathcal{P}}}\  {\rm s.t.}\ p_{i}\!=\!q_{i},\ i\!\in\!{{\mathcal{P}}},\ p_{i'}\neq q_{i'},\ i'\in {\mathcal{N}}\backslash{{\mathcal{P}}},&
	\label{ETL}
	\end{flalign}	
	where $w_{{\mathcal{P}}}$ is defined for each ${\mathcal{P}}\subseteq{\mathcal{N}}$ (which is equivalent to ${\mathcal{P}}\in{\bar {\mathcal{N}}}$).
	The above equation implies that, for a specific set of dimensions ${\mathcal{P}}$, the elements of $\tW$, which satisfy that the $N$-dimensional coordinate ${\bf p}$ are the same as the coordinate ${\bf q}$ on dimensions only in ${\mathcal{P}}$, share the same weight $w_{{\mathcal{P}}}$. Due to $|{\mathcal{N}}|=N$, there are $2^N$ different elements $w_{{\mathcal{P}}}$ in $\tW$. Although this pattern is intricate, we will proceed to demonstrate its equivalence to our expression. The $p_1,...,p_N$-th element of $\tY$ is given by
	\begin{align}
	& y_{p_1,...,p_N}\nonumber\\
	&=\sum_{q_N=1}^{M_N}\sum_{q_{N-1}=1}^{M_{N-1}}\cdots\sum_{q_{1}=1}^{M_{1}}{\bf W}_{[{\bf p}, (q_1, q_2,...,q_N)^T]}\cdot x_{q_1,...,q_N}\nonumber\\
	&=\sum_{{\mathcal{P}}\subseteq {\mathcal{N}}}w_{{\mathcal{P}}}\sum_{q_{i}=p_{i},i\in{\mathcal{P}} \atop q_{i'}\neq p_{i'}, i' \in {\mathcal{N}}\backslash {\mathcal{P}} } x_{q_1,..,q_N}\nonumber\\
	&= w_{\emptyset}\!\!\!\!\sum_{q_1,...,q_N}\!\!\!\!x_{q_1,q_2,..,q_N} + (w_{\{1\}}\!-\! w_{\emptyset})\!\!\!\!\sum_{q_2,...,q_N}\!\!\!\!x_{p_1,q_2,..,q_N} +\cdots \nonumber\\
	&\ \ +(w_{\{N\}}\!-\! w_{\emptyset})\!\!\!\!\sum_{q_1,...,q_{N-1}}\!\!\!\!x_{q_1,..,q_{N-1},p_N}\nonumber\\
	& \ \ 
	+ \![w_{\{1,2\}}\!-\!(w_{\{1\}}\!-\!w_{\emptyset})\!-\!(w_{\{2\}}\!-\!w_{\emptyset})\!-\!w_{\emptyset}]\!\!\!\!\sum_{q_3,...,q_N}\!\!\!x_{p_1,p_2,q_3,...,q_N}\nonumber\\
	& \ \ +\cdots\nonumber\\
	&\ \ +(w_{\{1,2,...,N\}}-\cdots-w_{\emptyset}) x_{p_1,...,p_N}\nonumber\\
	&= \sum_{{\mathcal{P}}\subseteq {\mathcal{N}}}{\hat w}_{{\mathcal{P}}}\sum_{q_{i}=p_{i},i\in{\mathcal{P}} \atop q_{i'}, i' \in {\mathcal{N}}\backslash {\mathcal{P}} } x_{q_1,..,q_N},
	\end{align}
    where ${\hat w}_{\emptyset}=w_{\emptyset}$ and ${\hat w}_{{\mathcal{P}}}=w_{{\mathcal{P}}}-\sum_{{\mathcal U}\subset {\mathcal{P}}}{\hat w}_{{\mathcal U}}$. We use ${\hat \tX}_{{\mathcal A}}\in{\mathbb R}^{M_1\times \cdots \times M_N},{\mathcal A}\subseteq{\mathcal{N}}$ to represent the tensor obtained by applying the summation operation over the dimensions ${\mathcal A}$ of tensor $\tX$, which is repeated over the dimensions ${\mathcal A}$ to match the original shape. Note that ${\hat \tX}_{\emptyset}=\tX$. A single term in the above formula can be represented as follows
	\begin{align}
		\begin{split}
		{\hat w}_{{\mathcal{P}}}\sum_{q_{i}=p_{i},i\in{\mathcal{P}} \atop q_i', i' \in {\mathcal{N}}\backslash {\mathcal{P}} } \!\!\! x_{q_1,..,q_N} = {\hat w}_{{\mathcal{P}}}{\hat \tX}_{{\mathcal{N}}\backslash{\mathcal{P}}[p_1,p_2,...,p_N]},
		\end{split}
	\end{align} 
	Based on the above formula, we have $y_{p_1,...,p_N}=\sum_{{\mathcal{P}}\subseteq {\mathcal{N}}}{\hat w}_{{\mathcal{P}}}{\hat \tX}_{{\mathcal{N}}\backslash{\mathcal{P}}[p_1,p_2,...,p_N]}$. Thus, ${\rm FC}_{\rm {MDE}}(\tX)$ can be expressed as 
	\begin{align}
	\begin{split}
		&\textstyle{\rm FC}_{\rm {MDE}}(\tX)  = \sum_{{\mathcal{P}}\subseteq {\mathcal{N}}}{\hat w}_{{\mathcal{P}}}{\hat \tX}_{{\mathcal{N}}\backslash{\mathcal{P}}}=\sum_{{\mathcal{P}}\subseteq {\mathcal{N}}}{\hat w}_{{\mathcal{N}}\backslash{\mathcal{P}}}{\hat \tX}_{{\mathcal{P}}}\\
		&\quad=\sum_{{\mathcal{P}}\subseteq {\mathcal{N}}}(\Pi_{n\in{\mathcal{P}}}M_n)\cdot{\hat w}_{{\mathcal{N}}\backslash{\mathcal{P}}}{\bar \tX}_{{\mathcal{P}}}=\sum_{{\mathcal{P}}\subseteq {\mathcal{N}}}{\bar w}_{{\mathcal{P}}}{\bar \tX}_{{\mathcal{P}}},
	\end{split}
	\end{align}
	where ${\bar w}_{{\mathcal{P}}} = (\Pi_{n\in{\mathcal{P}}}M_n)\cdot{\hat w}_{{\mathcal{N}}\backslash{\mathcal{P}}}$. 
	
	Subsequently, we consider the case where bias exists. We reshape the bias to $\tB\in\mathbb{R}^{M_1\times\cdots\times M_N}$. When the elements in $\tX$ are all zero, \eqref{multidimensional equivariance} degenerates to
	\begin{align}
	\textstyle{	\tB=\pi_{M_n}\circ_{n}\tB,\ \forall {\pi}_{M_n}\in {\mathbb{S}}_{M_n},\ \forall n\in\mathcal{N},}
	\end{align}
	which implies that $\tB=b{\bf 1}$. Therefore, ${\rm FC}_{\rm {MDE}}(\tX)$ can be formulated as $\textstyle{{\rm FC}_{\rm {MDE}}(\tX) = \sum_{{\mathcal{P}}\subseteq {\mathcal{N}}}{\bar w}_{{\mathcal{P}}}{\bar \tX}_{{\mathcal{P}}} + b{\bf 1}}$. 
	This expression can be readily extended to scenarios where $D_{\rm I}>1$ and $D_{\rm O}>1$.
	This completes the proof.
	
\section{Definitions of Tensors and Sets in ${\rm FC}_{\rm HOE}$}\label{HOE detail}

The indexing of specific features of the input and output tensors can be represented as $\tX_{[i_1,...,i_p, :]}$ and ${\tilde \tX}_{[o_1,...,o_q, :]}$, respectively. On this basis, we define ${\tilde {\mathcal{N}}}_{p\!-\!q}$ as the set of all non-repetitive partitions of the index set $\{o_1,...,o_q, i_1,...,i_p\}$. The number of ways to partition a set with cardinality $n$ is $B(n)$, where $B(\cdot)$ represents the Bell number, meaning that $|{\tilde {\mathcal{N}}}_{p\!-\!q}| = B(p+q)$. We suppose ${\mathcal P}\in {\tilde {\mathcal{N}}}_{p\!-\!q}$, and the set can be expressed as 
\begin{align}
	{\mathcal P} = \{{\mathcal O}_1,...,{\mathcal O}_{N_{O}}, {\mathcal T}_1,...,{\mathcal T}_{N_{T}}, {\mathcal I}_1,...,{\mathcal I}_{N_{I}}\},
\end{align} 
where ${\mathcal I}$ only contains the elements of $i$, ${\mathcal O}$ only contains the elements of $o$, and ${\mathcal T}$ contains both $i$ and $o$. 
Taking ${\tilde {\mathcal{N}}}_{1\!-\!2}$ as an example, it contains the following sets
\begin{flalign}
	&{\mathcal P}_1 \!=\! \{o_1,o_2,i_1\}\!\Rightarrow\! {\mathcal T}_1 \!=\! \{o_1, o_2, i_1\},&
	\\
	&{\mathcal P}_2 \!\!=\!\! \{\!\{o_1\},\!\{o_2\},\!\{i_1\}\!\} \!\Rightarrow\! {\mathcal O}_1 \!\!=\!\! \{o_1\!\}, {\mathcal O}_2 \!\!=\!\! \{o_2\}, {\mathcal I}_1 \!\!=\!\! \{i_1\!\},&\\
	&{\mathcal P}_3 \!=\! \{\{o_1\},\{o_2, i_1\}\} \!\Rightarrow\! {\mathcal O}_1 \!\!=\!\! \{o_1\}, {\mathcal T}_1 \!\!=\!\! \{o_2, i_1\},&\\
	&{\mathcal P}_4 \!=\! \{\{o_2\},\{o_1, i_1\}\} \!\Rightarrow\!{\mathcal O}_1 \!=\! \{o_2\}, {\mathcal T}_1 = \{o_1, i_1\},&\\
	&{\mathcal P}_5 \!=\! \{\{o_1, o_2\},\{i_1\}\} \!\Rightarrow\! {\mathcal O}_1 \!=\! \{o_1, o_2\}, {\mathcal I}_1 = \{i_1\}.&
\end{flalign}

\begin{figure}[htbp]
	\centering
	\includegraphics[width=0.45\textwidth]{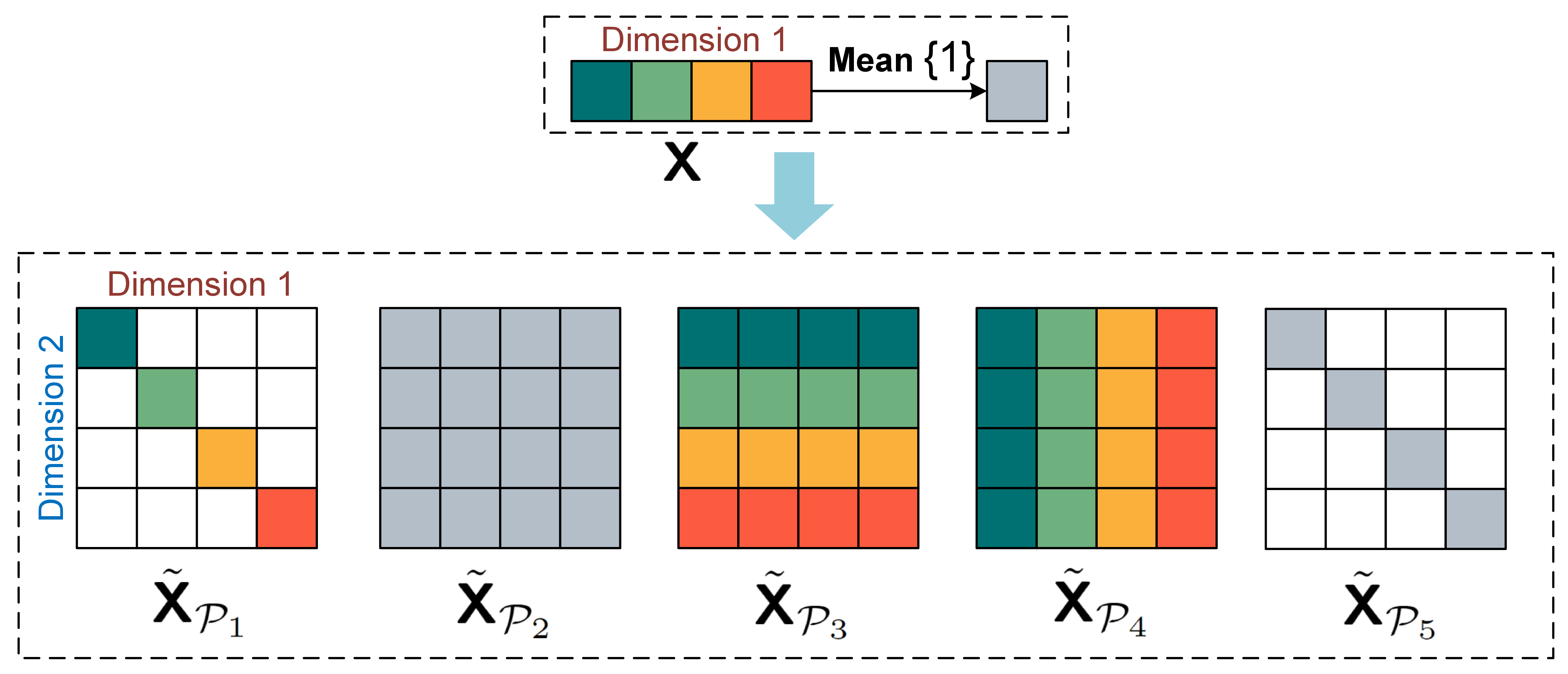}
	\caption{The construction method for ${\tilde \tX}_{{\mathcal P}}$ of the $1$-$2$-order equivariant module.}
	\label{1-2-order tensors}
\end{figure}

Each ${\mathcal P}$ corresponds to a tensor ${\tilde \tX}_{\mathcal P}$ with a certain pattern. To better describes the pattern, we define a mapping $\psi(\cdot)$ based on each set ${\mathcal P}$, as follows:
\begin{align}
	\psi(o_s) = 
	\begin{cases}
		{\tilde o}_{n},\  o_s\in {\mathcal O}_n\\
		t_n, \ o_s\in {\mathcal T}_n
	\end{cases}
	,\ 		
	\psi(i_s) = 
	\begin{cases}
		{\tilde i}_{n},\  i_s\in {\mathcal I}_n\\
		t_n, \ i_s\in {\mathcal T}_n
	\end{cases}
\end{align}
The values of matrices ${\tilde \tX}_{\mathcal P}$ and $\tX$ are connected through the set ${\mathcal P}$ and mapping $\psi(\cdot)$, as follows
\begin{align}
	&{\tilde \tX}_{{\mathcal P}[\psi(o_1), \psi(o_2), ..., \psi(o_q), :]} \\
	&\!= \frac{1}{M^{N_I}}\sum\limits_{{\tilde i}_{1}=1}^{M}\cdots \sum\limits_{{\tilde i}_{N_{I}}=1}^{M}\tX_{[\psi(i_1), \psi(i_2), ..., \psi(i_q), :]},\ \forall {\tilde o}_{n}, t_n\in\mathcal{M},\nonumber
\end{align}
where $\mathcal{M} = \{1,...,M\}$. For the indexes that can not be mapped from $\psi(\cdot)$, the elements of ${\tilde \tX}$ are set to be zeros, i.e., ${\tilde \tX}_{{\mathcal P}[o_1, o_2, ..., o_q]} = 0$. Taking ${\tilde {\mathcal{N}}}_{1\!-\!2}$ as an example, we have the following output tensors:
\begin{align}
	&{\mathcal P}_1:\psi(o_1)=\psi(o_2)=\psi(i_1)=t_1,\\
	&{\mathcal P}_2:\psi(o_1)={\tilde o}_1, \psi(o_2)={\tilde o}_2, \psi(i_1)={\tilde i}_1,\\
	&{\mathcal P}_3:\psi(o_1)={\tilde o}_1, \psi(o_2)=\psi(i_1)=t_1,\\
	&{\mathcal P}_4:\psi(o_2)={\tilde o}_1, \psi(o_1)=\psi(i_1)=t_1,\\
	&{\mathcal P}_5:\psi(o_1)=\psi(o_2)={\tilde o}_1, \psi(i_1)={\tilde i}_1,\\
	&\qquad\qquad\qquad\qquad\Downarrow\nonumber\\
	&{\tilde \tX}_{{\mathcal P}_1[t_1, t_1, :]} =  \tX_{[t_1, :]},\ \forall t_1\in{\mathcal{M}},\\
	&{\tilde \tX}_{{\mathcal P}_2[{\tilde o}_1, {\tilde o}_2, :]} = \frac{1}{M}\sum\nolimits_{{\tilde i}_1=1}^{M}\tX_{[{\tilde i}_1, :]},\ \forall {\tilde o}_1, {\tilde o}_2\in{\mathcal{M}},\\
	&{\tilde \tX}_{{\mathcal P}_3[{\tilde o}_1, t_1, :]} = \tX_{[t_1, :]},\ \forall {\tilde o}_1, t_1\in{\mathcal{M}},\\
	&{\tilde \tX}_{{\mathcal P}_4[t_1, {\tilde o}_1, :]} = \tX_{[t_1, :]},\ \forall {\tilde o}_1, t_1 \in{\mathcal{M}},\\
	&{\tilde \tX}_{{\mathcal P}_5[{\tilde o}_1, {\tilde o}_1, :]} = \frac{1}{M}\sum\nolimits_{{\tilde i}_1=1}^{M}\tX_{[{\tilde i}_1, :]},\ \forall {\tilde o}_1\in{\mathcal{M}},
\end{align}
Based on the above expressions, we present the construction method for the five output tensors of the $1$-$2$-order equivariant linear layer as \figref{1-2-order tensors}, where each small box represents a feature with length $D_{\rm I}$, and the blank boxes represent features with all-zero elements.

\begin{figure}[htbp]
	\centering
	\includegraphics[width=0.4\textwidth]{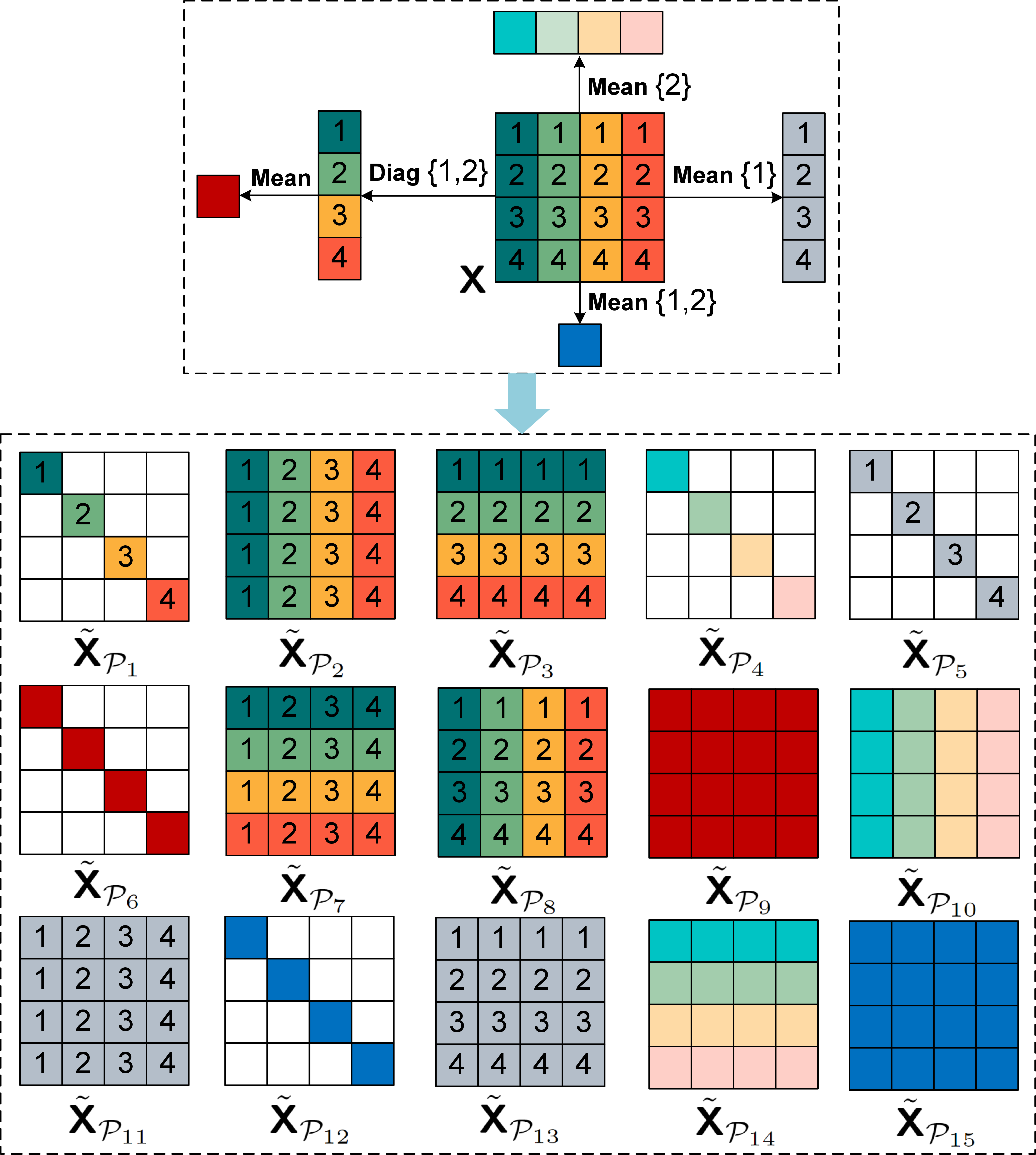}
	\caption{The construction method for the output tensors of the $2$-$2$-order equivariant layer.}
	\label{2-2-order tensors}
\end{figure}

Furthermore, by following the steps outlined in this appendix, we present the tensors of the 2-2 order equivariant layer in \figref{2-2-order tensors}. In this figure, the rows and columns of the input tensor are indicated by numbers and colors, respectively. The operation ${\rm Diag}\{1, 2\}$ represents selecting the diagonal elements from the first and second dimensions of the tensor.

Building on the aforementioned representation of weights, we next define the representation of bias tensor $\tB_{\mathcal Q}$. We define ${\tilde {\mathcal{N}}}_{q} = \{o_1,...,o_q\}$. Then, for ${\mathcal Q}\in {\tilde {\mathcal{N}}}_{q}$, we have 
\begin{align}
	\tB_{{\mathcal Q}[\psi(o_1), \psi(o_2), ..., \psi(o_q), :]} = {\bf b}_{{\mathcal Q}},\ \forall {\tilde o}_{n}\in\mathcal{M},
\end{align}
where ${\bf b}_{{\mathcal Q}}\in\mathbb{R}^{D_{\rm O}\times 1}$ is the learnable bias.
For $1$-$2$-order and $2$-$2$-order equivariant layers, we have $q=2$, which leads to
\begin{align}
	&\tB_{{\mathcal Q}_1[{\tilde o}_1, {\tilde o}_1, :]} =  {\bf b}_{{\mathcal Q}_1},\ \forall {\tilde o}_1\in{\mathcal{M}},\\
	&\tB_{{\mathcal Q}_2[{\tilde o}_1, {\tilde o}_2, :]} =  {\bf b}_{{\mathcal Q}_2},\ \forall {\tilde o}_1, {\tilde o}_2\in{\mathcal{M}}.
\end{align}

\section{Proof of Proposition \ref{ppn stack}}\label{pf stack}
Proposition \ref{ppn stack} can be expressed as the following two points:
\begin{itemize}
	\item If $g_1:\mathbb{R}^{\scriptsize{\overbrace{M\times \cdots\times M}^{p}}\times D_X}\to \mathbb{R}^{\scriptsize{\overbrace{M\times \cdots\times M}^{p}}\times D_{Y'}}$ satisfies $p$-dimensional equivariance, $g_2:\mathbb{R}^{\scriptsize{\overbrace{M\times \cdots\times M}^{p}}\times D_{Y'}}\to \mathbb{R}^{\scriptsize{\overbrace{M\times \cdots\times M}^{q}}\times D_Y}$ satisfies $p$-$q$-order equivariance, then $f=g_2\circ g_1$ also satisfies $p$-$q$-order equivariance.
	\item If $h_1:\mathbb{R}^{M_1\times \cdots\times M_N\times D_X}\to \mathbb{R}^{M_1\times \cdots\times M_N\times D_Y}$ satisfies $N$-dimensional equivariance, $h_2:\mathbb{R}^{M_1\times \cdots\times M_N\times D_X}\to \mathbb{R}^{D_Y}$ satisfies $N$-dimensional invariance, then $f=h_2\circ h_1$ also satisfies $N$-dimensional invariance.
\end{itemize}

For the first point, we have
\begin{align}
	\begin{split}
		&f({\pi_{M}}\circ_{[1,...,p]}\tX)=g_2\left(g_1\left({\pi_{M}}\circ_{[1,...,p]}\tX\right)\right)\\
		&=g_2\left({\pi_{M}}\circ_{[1,...,p]}g_1\left(\tX\right)\right) = {\pi_{M}}\circ_{[1,...,q]}g_2\left(g_1\left(\tX\right)\right)\\
		&={\pi_{M}}\circ_{[1,...,q]}f(\tX),\ \forall {\pi_{M}}\in {\mathbb{S}}_{M}.
	\end{split}
\end{align}
For the second point, we have
\begin{align}
	&f({\pi_{M_n}}\circ_{n}\tX)=h_2\left(h_1\left({\pi_{M_n}}\!\circ_{n}\tX\right)\right)=h_2\left({\pi_{M_n}}\circ_{n}h_1\left(\tX\right)\right) \notag \\
	&= h_2\left(h_1\left(\tX\right)\right)=f(\tX),\ \forall {\pi_{M_n}}\in {\mathbb{S}}_{M_n}, \forall n\in\mathcal{N}.
\end{align}
This completes the proof.

\section{Case of Multiple Optimal Solutions}

In this appendix, we discuss the research paradigm proposed in Sections \ref{precoding design sec} and \ref{US design sec}, which involves exploiting TE of mappings from available information to optimal solution in optimization problem, under the case where the optimization problem has multiple optimal solutions.

The theory in TE framework that exploiting TE within the target mapping accommodates both single optimal solutions and cases with multiple optimal solutions. Specifically, if the original problem has multiple optimal solutions, the optimization problem based on the permuted available information (specifically, the CSI in optimization problem of precoding) will also exhibit the same number of optimal solutions, which correspond one-to-one with the optimal solutions of the original problem through the same permutation. Thus, the mapping from available information to a certain optimal solution still satisfies TE.

Taking the problem of sum rate maximization in precoding design as an example.
\begin{align}
	\begin{split}
		& \max_{\tW}\ \sum_{k=1}^KR_k(\tH,\tW,\sigma^2)\ \ {\rm s.t.}\ \sum_{k=1}^{K}{\rm Tr}\left({\bf W}_k{\bf W}^H_k\right)\leq P_{\rm T}.
	\end{split}
	\label{R3 3D precoding problem}
\end{align} 
To simplify the description, we take the single-dimensional equivariance as an example. According to the derivation in \textbf{Appendix A}, we have the following proposition
\begin{ppn}
	The objective function of  $\langle\tW,\{\tH,\sigma^2\}\rangle_{\rm P}$ is equal to those of $\langle\pi_{K}\circ_1\tW,\{\pi_{K}\circ_1\tH,\sigma^2\}\rangle_{\rm P}$, for all $\pi_{K}\in{\mathbb{S}}_K$. Specifically, if $\langle\tW^{\star},\{\tH,\sigma^2\}\rangle_{\rm P}$ achieves the optimal objective function, then $\langle\pi_{K}\circ_1\tW^{\star},\{\pi_{K}\circ_1\tH,\sigma^2\}\rangle_{\rm P}$ can also achieve their optimal objective functions.
	\label{R3 precoding 3D PE ppn}
\end{ppn}
The above proposition illustrates that the optimal solutions of the optimization problem based on $\tH,\sigma^2$ and the optimization problem based on $\pi_{K}\circ_1\tH,\sigma^2$, $\forall \pi_{K}\in{\mathbb{S}}_K$, have a one-to-one correspondence. Therefore, the following conclusion can be derived.
\begin{ppn}
	For precoding design problem \eqref{R3 3D precoding problem}, the number of optimal solutions for the problem based on CSI $\tH,\sigma^2$ is the same as that for the optimization problem based on CSI $\pi_{K}\circ_1\tH,\sigma^2$, $\forall \pi_{K}\in{\mathbb{S}}_K$.
	\label{R3 same solution number}
\end{ppn}
If the above optimization problem has $T$ optimal solutions, according to \ppnref{R3 precoding 3D PE ppn} and \ppnref{R3 same solution number}, they have the following relationship: for the original problem
\begin{align}
	&G_{{\rm P}(1)}(\tH, \sigma^2)=\tW^{\star}_{(1)},\\
	&G_{{\rm P}(2)}(\tH, \sigma^2)=\tW^{\star}_{(2)}\\
	&\qquad\qquad\cdots\nonumber\\
	&G_{{\rm P}(T)}(\tH, \sigma^2)=\tW^{\star}_{(T)}.
\end{align}
For the problem based on permuted CSI
\begin{align}
	&G_{{\rm P}(1)}(\pi_{K}\circ_1\tH, \sigma^2)=\pi_{K}\circ_1\tW^{\star}_{(1)},\ \forall \pi_{K}\in {\mathbb{S}}_K,\\
	&G_{{\rm P}(2)}(\pi_{K}\circ_1\tH, \sigma^2)=\pi_{K}\circ_1\tW^{\star}_{(2)},\ \forall \pi_{K}\in {\mathbb{S}}_K,\\
	&\qquad\qquad\qquad\qquad\cdots\nonumber\\
	&G_{{\rm P}(T)}(\pi_{K}\circ_1\tH, \sigma^2)=\pi_{K}\circ_1\tW^{\star}_{(T)},\ \forall \pi_{K}\in {\mathbb{S}}_K,
\end{align}
where $\tW^{\star}_{(1)},...,\tW^{\star}_{(T)}$ are the $T$ optimal solutions of the optimization problem based on CSI $\tH,\sigma^2$. $\pi_{K}\circ_1\tW^{\star}_{(1)},...,\pi_{K}\circ_1\tW^{\star}_{(T)}$ are the $T$ optimal solutions of the optimization problem based on CSI $\pi_{K}\circ_1\tH,\sigma^2$, $\forall \pi_{K}\in{\mathbb{S}}_K$. $G_{{\rm P}(1)},...,G_{{\rm P}(T)}$ are the mappings from the available CSI to the $T$ optimal solutions, respectively. This indicates that even there are multiple optimal solutions, the mapping to obtain a specific optimal solution still satisfies the equivariance.

\begin{figure}[htbp]
	\centering
	\begin{minipage}[b]{0.45\textwidth}
		\centering
		\includegraphics[width=\textwidth]{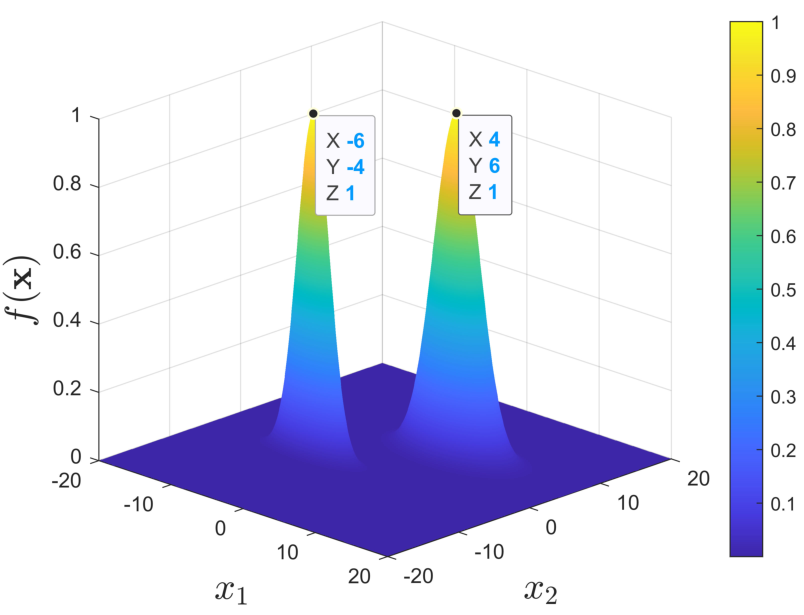}
		\caption{Objective function with ${\bf A}$.}
		\label{R3 toy_fun}
	\end{minipage}
	\hspace{0.05\textwidth} 
	\begin{minipage}[b]{0.45\textwidth}
		\centering
		\includegraphics[width=\textwidth]{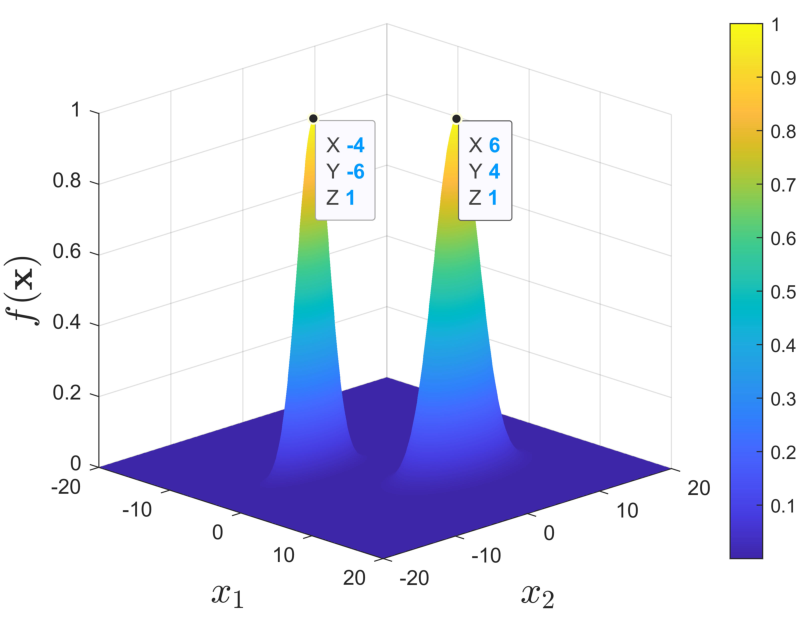}
		\caption{Objective function with ${\bf A}'$.}
		\label{R3 toy_fun_permute}
	\end{minipage}
\end{figure}

Furthermore, we take the optimization problem $\max\limits_{{\bf x}} f({\bf x})$ as an example, where 
\begin{align}
	\begin{split}
		&f({\bf x})=\max \left\{{\rm e}^{-\left( \frac{x}{a_{2,2}} - a_{1,1} \right)^2 - \left( \frac{y}{a_{1,2}} - a_{2,1} \right)^2}\right., \\
		&\qquad\qquad\qquad\qquad \left.{\rm e}^{ -\left( \frac{x}{a_{2, 1}} + a_{1, 2} \right)^2 - \left( \frac{y}{a_{1, 1}} + a_{2, 2} \right)^2}\right\}.
	\end{split}
\end{align}
To simplify the description, we denote ${\bf A} = \begin{bmatrix}
	a_{1, 1} & a_{1, 2}\\
	a_{2, 1} & a_{2, 2}
\end{bmatrix}$. With ${\bf A} = \begin{bmatrix}
	1 & 2\\
	3 & 4
\end{bmatrix}$
, the objective function is shown in \figref{R3 toy_fun}, with two optimal solutions denoted as ${\bf x}^{\star}_{(1)} = 
\begin{bmatrix}
	a_{1,1}*a_{2, 2} \\
	a_{2,1}*a_{1, 2}
\end{bmatrix} = 
\begin{bmatrix}
	4 \\
	6
\end{bmatrix}
$ and ${\bf x}^{\star}_{(2)} = 
\begin{bmatrix}
	-a_{1,2}*a_{2, 1} \\
	-a_{2,2}*a_{1, 1}
\end{bmatrix} = 
\begin{bmatrix}
	-6 \\
	-4
\end{bmatrix}$. As shown in \figref{R3 toy_fun_permute}, when we apply a permutation to the first dimension of ${\bf A}$, for instance, let ${\bf A}' = \pi \circ_{1} {\bf A} =  \begin{bmatrix}
	3 & 4\\
	1 & 2
\end{bmatrix}$, the corresponding optimal solutions become ${\bf x}^{'\star}_{(1)} = \begin{bmatrix}
	6 \\
	4
\end{bmatrix}$ and ${\bf x}^{'\star}_{(2)}=\begin{bmatrix}
	-4 \\
	-6
\end{bmatrix}$, where ${\bf x}^{'\star}_{(1)} = \pi \circ_{1} {\bf x}^{\star}_{(1)}$and ${\bf x}^{'\star}_{(2)} = \pi \circ_{1} {\bf x}^{\star}_{(2)}$.

Let $G_{f(1)}(\cdot)$ and $G_{f(2)}(\cdot)$ represent the mappings from ${\bf A}$ to ${\bf x}^{\star}_{(1)}$ and ${\bf x}^{\star}_{(2)}$, respectively, then the above process can be expressed as
\begin{align}
	G_{f(1)}({\bf A}) = {\bf x}^{\star}_{(1)},\ G_{f(1)}(\pi\circ_1{\bf A}) = \pi\circ_1{\bf x}^{\star}_{(1)},\\
	G_{f(2)}({\bf A}) = {\bf x}^{\star}_{(2)},\ G_{f(2)}(\pi\circ_1{\bf A}) = \pi\circ_1{\bf x}^{\star}_{(2)}.
\end{align}
The above equation verifies that the number of optimal solutions for these two optimization problems is the same, and that the mappings to their respective specific optimal solutions, such as $G_{f(1)}$ and $G_{f(2)}$, both satisfy the property of equivariance.

Another issue that needs attention is the potential impact of multiple optimal solutions on performance when using equivariant networks to fit the mapping to the optimal solution. Taking the above issue as an example, directly using ${\bf A}$ as input to the equivariant NN to train for ${\bf x}^{\star}$ may cause the network to simultaneously fit both mappings $G_{f(1)}(\cdot)$ and $G_{f(2)}(\cdot)$, leading to suboptimal performance. As pointed in \cite{sun2018learning}, when a specific initial value (e.g., ${\bf x}_0$) is provided as input, the network can converge to the corresponding unique optimal solution during training, thereby mitigating the performance degradation caused by multiple optimal solutions. This implies that when designing the TE network, incorporating a specific initial solution as part of the input could be a potential approach to avoid performance degradation caused by multiple optimal solutions, which can be achieved through operations such as concatenation.

\section{Low-Complexity Pattern Analysis}
In this appendix, we provide the theoretical analysis underpinning the complexity reduction approach proposed in Section \ref{PE linear layer sec}, along with the final method proposed based on this analysis.

This method for reducing computational complexity and parameter count is applicable to both the multidimensional equivariant module and the high-order equivariant module. Here, we take the multidimensional equivariant module as an example, which can be rewritten as:
\begin{align}
	{\rm FC}_{\rm MDE}(\tX) = \sum_{{\mathcal P}\in {\bar {\mathcal N}}}\left({\bar \tX}_{\mathcal P}\times\!{\bf W}_{{\mathcal{P}}}\right)+{\bf 1}\!\otimes_N\!{\bf b}^T_{\rm MDE},
	\label{R4 MDE Layer RS2}
\end{align}
where ${\bar {\mathcal N}}$ denote the set containing all subsets of ${\mathcal N} = \{1,...,N\}$ (power set). The computational complexity of the module is related to the cardinality of set ${\bar {\mathcal N}}$ by the relationship $\mathcal{O}(|{\bar {\mathcal N}}|)$, where $|{\bar {\mathcal N}}|=2^N$. As mentioned in our initial paper, extracting specific submatrices from the $2^N$ matrices of ${\bf W}_{{\mathcal P}}$ can reduce the computational complexity of this module, and it is easy to prove that this operation can still maintain the multidimensional equivariance.
Considering that ${\mathcal P}\in {\bar {\mathcal N}}$ in \eqref{R4 MDE Layer RS2}, the method for \textit{reducing complexity essentially means selecting an appropriate subset ${\bar {\mathcal N}}'$ to replace ${\bar {\mathcal N}}$}. 

When ${\mathcal P}=\varnothing$, ${\bar \tX}_{{\mathcal P}}=\tX$ represents the tensor itself. When ${\mathcal P}$ is a non-empty set, ${\bar \tX}_{{\mathcal P}}$ is the mean of $\tX$ along the dimensions included in ${\mathcal P}$, characterizing the global feature of $\tX$ in the dimensions encompassed by ${\mathcal P}$. This indicates that the target of \eqref{R4 MDE Layer RS2} is to perform linear interactions of tensor itself and the global feature from all the combinations of its dimensions. 

We can categorize the global feature $\{{\bar \tX}_{{\mathcal P}}\}_{{\mathcal P}\in{\bar {\mathcal N}}, {\mathcal P}\neq\varnothing}$ into two types: one is the global feature ${\bar \tX}_{{\mathcal P}}$ on a single dimension ($|{\mathcal P}|=1$), and the other is the global feature ${\bar \tX}_{{\mathcal P}}$ across multiple dimensions ($|{\mathcal P}|>1$). \textit{If a module only considers features on a single dimension while ignoring features on multiple dimensions, can it be expected to achieve good performance? The answer is no for a single module, but maybe yes for a network composed of multiple stacked modules.}

We analyze using two-dimensional equivariance ($N=2$) and $D_I=D_O=1$ as an example, which can be generalized to any higher dimensions. For a single module, ${\bar {\mathcal N}}=\{\varnothing, \{1\}, \{2\}, \{1, 2\}\}$. Accordingly, the operation of the two-dimensional equivariant module is given by
\begin{align}
	&{\bar {\mathcal N}}=\{\varnothing, \{1\}, \{2\}, \{1, 2\}\}:\nonumber\\
	&{\bf Y} = w_{\varnothing}{\bf X} + w_{1}{\bar {\bf X}}_{\{1\}} + w_{2}{\bar {\bf X}}_{\{2\}} + w_{\{1,2\}}{\bar {\bf X}}_{\{1,2\}} +b\cdot{\bf 1},\\
	&y_{i,j} = w_{\varnothing}x_{i,j} + \frac{w_{1}}{M_1}\sum_{m=1}^{M_1}x_{m,j} + \frac{w_{2}}{M_2}\sum_{m=1}^{M_2}x_{i,m} \nonumber\\
	& \qquad\qquad \qquad\qquad +\frac{w_{\{1,2\}}}{M_1M_2}\sum_{m_1=1}^{M_1}\sum_{m_2=1}^{M_2}x_{m_1,m_2} +  b.
	\label{R4 2D layer}
\end{align}	
where ${\bf X}\in\mathbb{R}^{M_1\times M_2}$ is the input, ${\bf Y}\in\mathbb{R}^{M_1\times M_2}$ is the output, and $x_{i,j}$ and $y_{i,j}$ are the elements of ${\bf X}$ and ${\bf Y}$ at the $i$-th row and $j$-th column, respectively.
If we retain only the global features of each individual dimension, then the set ${\bar {\mathcal N}}$ is replaced by ${\bar {\mathcal N}}_1 = \{\varnothing, 1, 2\}$.	
\begin{align}
	&{\bar {\mathcal N}}_1=\{\varnothing, \{1\}, \{2\}\}:\nonumber\\
	&{\bf Y} = w_{\varnothing}{\bf X} + w_{1}{\bar {\bf X}}_{\{1\}} + w_{2}{\bar {\bf X}}_{\{2\}} + b\cdot{\bf 1},\\
	&y_{i,j} = w_{\varnothing}x_{i,j} + \frac{w_{1}}{M_1}\sum_{m=1}^{M_1}x_{m,j} + \frac{w_{2}}{M_2}\sum_{m=1}^{M_2}x_{i,m} + b.
	\label{R4 2D layer single dimension}
\end{align}
The output $y_{i,j}$ in Equation \eqref{R4 2D layer single dimension} contains the global features of the row and column where \( x \) is located. However, compared to the output in \eqref{R4 2D layer}, the output in Equation \eqref{R4 2D layer single dimension} lacks the global feature shared across dimensions 1 and 2, which includes features from other rows and columns, i.e., $x_{m,n}, m\neq i, n\neq j$. This indicates that even if the features of other rows and columns undergo significant changes, the output $y_{i,j}$ corresponding to $x_{i,j}$ will remain unchanged, which maybe unreasonable. For example, in the case of a MU-MIMO channel, it is clearly unreasonable that arbitrary changes in the channels of other users' other antennas do not affect the transmission scheme design of certain antenna from current user. This shows that \textit{in the case of a single layer, removing the global feature shared across multiple dimensions in the multidimensional equivariant module will cause the network to lose the ability to capture cross-dimensional global features.}

\begin{flalign}
	&\text{The first layer:}&\nonumber\\
	&y^{(1)}_{i,j} = w^{(1)}_{\varnothing}x_{i,j} + \frac{w^{(1)}_{1}}{M_1}\sum_{m=1}^{M_1}x_{m,j} + \frac{w^{(1)}_{2}}{M_2}\sum_{m=1}^{M_2}x_{i,m} + b^{(1)},&\label{R4 2D 2layers single dimension 1}\\
	&\text{The second layer:}&\nonumber\\
	&y^{(2)}_{i,j} = w^{(2)}_{\varnothing}y^{(1)}_{i,j} + \frac{w^{(2)}_{1}}{M_1}\sum_{m=1}^{M_1}y^{(1)}_{m,j} + \frac{w^{(2)}_{2}}{M_2}\sum_{m=1}^{M_2}y^{(1)}_{i,m} + b^{(2)}.&
	\label{R4 2D 2layers single dimension 2}
\end{flalign}
Then, we consider a two-layer neural network (for the sake of clarity, we ignore element-wise operations such as activation functions), as shown in equations \eqref{R4 2D 2layers single dimension 1} and \eqref{R4 2D 2layers single dimension 2}. As mentioned in the previous paragraph, $y^{(1)}_{i,j}$ does not capture cross-dimensional features; it only captures the features of the row and column where $x_{i,j}$ is located. However, during the processing of the second layer, $y^{(2)}_{i,j}$ includes the features from $y^{(1)}_{i,j}$ of the corresponding row and column, where these $y^{(1)}_{i,j}$ respectively capture the features of their corresponding row and column dimensions. This enables $y^{(2)}_{i,j}$ to capture both the single dimensional features in the row and column of $x_{i,j}$ as well as the features across the two dimensions. This mechanism is similar to the receptive field mechanism of convolutional layers \cite{luo2016understanding}: although a single convolutional layer has a limited receptive field, it can gradually expand its receptive field by stacking multiple convolutional layers. This illustrates that \textit{for a network composed of multiple stacked layers, even if a single layer only possesses the ability to capture global features across individual dimensions, the network still has the potential to capture cross-dimensional global features}.

Based on the analysis above, if we need to construct a network containing $L$ layers of equivariant modules, Where the $l$-th layer of the equivariant module uses the set ${\bar {\mathcal N}}'_l$ to replace ${\bar {\mathcal N}}$ in \eqref{R4 MDE Layer RS2}. We propose the following three patterns for selecting matrices (i.e., constructing subset of ${\bar {\mathcal N}}$):
\begin{itemize}
	\item \textbf{Pattern 1:} The set ${\bar {\mathcal N}}'_l$ for each layer of the network is the same, which is 
	\begin{align}
		{\bar {\mathcal N}}^{\rm P1}_{l}=\{\varnothing, \{1\},\{2\},...,\{N\}\},\ \forall l.
	\end{align}
	When $N=3$, we have ${\bar {\mathcal N}}^{\rm P1}_{l}=\{\varnothing, \{1\},\{2\},\{3\}\}$. The computational complexity order of this module that related to $N$ has been reduced from ${\mathcal O}(2^N)$ to ${\mathcal O}(N+1)$.
	
	\item \textbf{Pattern 2:} The set ${\bar {\mathcal N}}'_{l}$ for each layer of the network can be different, but must include the empty set $\varnothing$. The included sets must satisfy $|{\mathcal P}|\leq 1$ and need to satisfy
	\begin{align}
		{\bar {\mathcal N}}^{\rm P2}_{1}\cup {\bar {\mathcal N}}^{\rm P2}_{2}\cup\cdots \cup{\bar {\mathcal N}}^{\rm P2}_{L}=\{\varnothing, \{1\},\{2\},...,\{N\}\}.
		\label{R4 pattern condition}
	\end{align}
	For example, when $N=3$ and $L=3$, the sets can be ${\bar {\mathcal N}}^{\rm P2}_{1} = \{\varnothing, \{1\}, \{2\}\}$,  ${\bar {\mathcal N}}^{\rm P2}_{2} = \{\varnothing, \{2\}, \{3\}\}$, and  ${\bar {\mathcal N}}^{\rm P2}_{3} = \{\varnothing, \{1\}, \{3\}\}$. The above equation ensures that there is no loss of global information capture for certain dimensions or combinations of dimensions throughout the entire network. The complexity order of this module that related to $N$ has been reduced from ${\mathcal O}(2^N)$ to ${\mathcal O}(N')$, where $N' = \frac{\sum_{l=1}^{L}|{\bar {\mathcal N}}^{\rm P2}_l|}{L}\leq N+1$.
	
	\item \textbf{Pattern 3:} The sets ${\bar {\mathcal N}}'_{l}$ of different layers can be different, but they only contain two elements, $\varnothing$ and ${\mathcal P}_l$, where $|{\mathcal P}_l|=1$, and they need to satisfy
	\begin{align}
		{\mathcal P}_{1}\cup {\mathcal P}_{2}\cup\cdots \cup{\mathcal P}_{L} =\{\{1\},\{2\},...,\{N\}\}.
	\end{align}
	For example, when $N=3$ and $L=3$, the sets can be ${\bar {\mathcal N}}^{\rm P3}_{1} = \{\varnothing, \{1\}\}$,  ${\bar {\mathcal N}}^{\rm P3}_{2} = \{\varnothing, \{2\}\}$, and  ${\bar {\mathcal N}}^{\rm P3}_{3} = \{\varnothing, \{3\}\}$. This pattern can be regarded as a special case of \textbf{Pattern 2}.
	The complexity order of this module that related to $N$ has been reduced from ${\mathcal O}(2^N)$ to ${\mathcal O}(2)$.
\end{itemize}

	\bibliographystyle{IEEEtran}
	\bibliography{Refabrv,IEEEfull}

\begin{thebibliography}{10}
\providecommand{\url}[1]{#1}
\csname url@samestyle\endcsname
\providecommand{\newblock}{\relax}
\providecommand{\bibinfo}[2]{#2}
\providecommand{\BIBentrySTDinterwordspacing}{\spaceskip=0pt\relax}
\providecommand{\BIBentryALTinterwordstretchfactor}{4}
\providecommand{\BIBentryALTinterwordspacing}{\spaceskip=\fontdimen2\font plus
\BIBentryALTinterwordstretchfactor\fontdimen3\font minus
  \fontdimen4\font\relax}
\providecommand{\BIBforeignlanguage}[2]{{%
\expandafter\ifx\csname l@#1\endcsname\relax
\typeout{** WARNING: IEEEtran.bst: No hyphenation pattern has been}%
\typeout{** loaded for the language `#1'. Using the pattern for}%
\typeout{** the default language instead.}%
\else
\language=\csname l@#1\endcsname
\fi
#2}}
\providecommand{\BIBdecl}{\relax}
\BIBdecl

\bibitem{6375940}
F.~Rusek, D.~Persson \emph{et~al.}, ``Scaling up {MIMO}: Opportunities and
  challenges with very large arrays,'' \emph{IEEE Signal Process. Mag.},
  vol.~30, no.~1, pp. 40--60, Jan. 2013.

\bibitem{wang2023road}
C.-X. Wang, X.~You \emph{et~al.}, ``On the road to 6{G}: {V}isions,
  requirements, key technologies and testbeds,'' \emph{IEEE Commun. Surv.
  Tutor.}, vol.~25, no.~2, pp. 905--974, Secondquarter 2023.

\bibitem{dimic2005downlink}
G.~Dimic and N.~D. Sidiropoulos, ``On downlink beamforming with greedy user
  selection: {P}erformance analysis and a simple new algorithm,'' \emph{IEEE
  Trans. Signal Process.}, vol.~53, no.~10, pp. 3857--3868, Sept. 2005.

\bibitem{4712693}
S.~S. Christensen, R.~Agarwal \emph{et~al.}, ``Weighted sum-rate maximization
  using weighted {MMSE} for {MIMO-BC} beamforming design,'' \emph{IEEE Trans.
  Wireless Commun.}, vol.~7, no.~12, pp. 4792--4799, Dec. 2008.

\bibitem{shi2011iteratively}
Q.~Shi, M.~Razaviyayn \emph{et~al.}, ``An iteratively weighted {MMSE} approach
  to distributed sum-utility maximization for a {MIMO} interfering broadcast
  channel,'' \emph{IEEE Trans. Signal Process.}, vol.~59, no.~9, pp.
  4331--4340, Sept. 2011.

\bibitem{zhao2023rethinking}
X.~Zhao, S.~Lu \emph{et~al.}, ``Rethinking {WMMSE}: Can its complexity scale
  linearly with the number of {BS }antennas?'' \emph{IEEE Trans. Signal
  Process.}, vol.~71, pp. 433--446, Feb. 2023.

\bibitem{1413598}
B.~Hochwald, C.~Peel, and A.~Swindlehurst, ``A vector-perturbation technique
  for near-capacity multiantenna multiuser communication-{P}art {II}:
  {P}erturbation,'' \emph{IEEE Trans. Commun.}, vol.~53, no.~3, pp. 537--544,
  2005.

\bibitem{1056659}
M.~Costa, ``Writing on dirty paper (corresp.),'' \emph{IEEE Trans. Inf.
  Theory}, vol.~29, no.~3, pp. 439--441, 1983.

\bibitem{8323218}
L.~Liu and W.~Yu, ``Massive connectivity with massive {MIMO}—{P}art {I}:
  {D}evice activity detection and channel estimation,'' \emph{IEEE Trans.
  Signal Process.}, vol.~66, no.~11, pp. 2933--2946, 2018.

\bibitem{hornik1989multilayer}
K.~Hornik, M.~Stinchcombe, and H.~White, ``Multilayer feedforward networks are
  universal approximators,'' \emph{Neural networks}, vol.~2, no.~5, pp.
  359--366, 1989.

\bibitem{yun2019transformers}
\BIBentryALTinterwordspacing
C.~Yun, S.~Bhojanapalli \emph{et~al.}, ``Are transformers universal
  approximators of sequence-to-sequence functions?'' \emph{arXiv preprint
  arXiv:1912.10077}, 2019. [Online]. Available:
  \url{https://arxiv.org/abs/1912.10077}
\BIBentrySTDinterwordspacing

\bibitem{letaief2019roadmap}
K.~B. Letaief, W.~Chen \emph{et~al.}, ``The roadmap to 6{G}: {AI} empowered
  wireless networks,'' \emph{IEEE Commun. Mag.}, vol.~57, no.~8, pp. 84--90,
  Aug. 2019.

\bibitem{kim2020deep}
J.~Kim, H.~Lee \emph{et~al.}, ``Deep learning methods for universal {MISO}
  beamforming,'' \emph{IEEE Wireless Commun. Lett.}, vol.~9, no.~11, pp.
  1894--1898, Nov. 2020.

\bibitem{huang2019fast}
H.~Huang, Y.~Peng \emph{et~al.}, ``Fast beamforming design via deep learning,''
  \emph{IEEE Trans. Veh. Technol.}, vol.~69, no.~1, pp. 1065--1069, Jan. 2019.

\bibitem{9322310}
S.~Lu, S.~Zhao, and Q.~Shi, ``Learning-based massive beamforming,'' in
  \emph{IEEE Glob. Commun. Conf., (GLOBECOM)}, Dec. 2020, pp. 1--6.

\bibitem{8935405}
W.~Xia, G.~Zheng \emph{et~al.}, ``A deep learning framework for optimization of
  {MISO} downlink beamforming,'' \emph{IEEE Trans. Commun.}, vol.~68, no.~3,
  pp. 1866--1880, Mar. 2020.

\bibitem{9516008}
J.~Shi, W.~Wang \emph{et~al.}, ``Deep learning-based robust precoding for
  massive {MIMO},'' \emph{IEEE Trans. Commun.}, vol.~69, no.~11, pp.
  7429--7443, Nov. 2021.

\bibitem{shi2023robust}
J.~Shi, A.-A. Lu \emph{et~al.}, ``Robust {WMMSE} precoder with deep learning
  design for massive {MIMO},'' \emph{IEEE Trans. Commun.}, vol.~71, no.~7, pp.
  3963--3976, Jul. 2023.

\bibitem{6832894}
E.~Björnson, M.~Bengtsson, and B.~Ottersten, ``Optimal multiuser transmit
  beamforming: A difficult problem with a simple solution structure [lecture
  notes],'' \emph{IEEE Signal Process. Mag.}, vol.~31, no.~4, pp. 142--148,
  Jul. 2014.

\bibitem{9667094}
L.~Pellaco, M.~Bengtsson, and J.~Jaldén, ``Matrix-inverse-free deep unfolding
  of the weighted {MMSE} beamforming algorithm,'' \emph{IEEE Open J. Commun.
  Soc.}, vol.~3, pp. 65--81, Dec. 2021.

\bibitem{9246287}
Q.~Hu, Y.~Cai \emph{et~al.}, ``Iterative algorithm induced deep-unfolding
  neural networks: Precoding design for multiuser {MIMO} systems,'' \emph{IEEE
  Trans. Wireless Commun.}, vol.~20, no.~2, pp. 1394--1410, Feb. 2021.

\bibitem{wang2024robust}
K.~Wang and A.~Liu, ``Robust {WMMSE}-based precoder with practice-oriented
  design for massive {MU-MIMO},'' \emph{IEEE Wireless Commun. Lett.}, vol.~13,
  no.~7, pp. 1858--1862, Jul. 2024.

\bibitem{8922744}
F.~Liang, C.~Shen \emph{et~al.}, ``Towards optimal power control via ensembling
  deep neural networks,'' \emph{IEEE Trans. Commun.}, vol.~68, no.~3, pp.
  1760--1776, Mar. 2020.

\bibitem{li2021user}
Y.~Li, S.~Han, and C.~Yang, ``User scheduling for uplink {OFDMA} systems by
  deep learning,'' in \emph{IEEE Wireless Commun. Networking Conf. (WCNC)},
  Nanjing, China, Mar. 2020, pp. 1--6.

\bibitem{xie2024learning}
B.~Xie, S.~Chen \emph{et~al.}, ``Learning-assisted user scheduling and
  beamforming for mm{W}ave vehicular networks,'' \emph{IEEE Trans. Veh.
  Technol.}, Early Access, 2024.

\bibitem{sun2018learning}
H.~Sun, X.~Chen \emph{et~al.}, ``Learning to optimize: {T}raining deep neural
  networks for interference management,'' \emph{IEEE Trans. Signal Process.},
  vol.~66, no.~20, pp. 5438--5453, Oct. 2018.

\bibitem{9674231}
X.~Yi, ``Asymptotic spectral representation of linear convolutional layers,''
  \emph{IEEE Trans. Signal Process.}, vol.~70, pp. 566--581, Jan. 2022.

\bibitem{zaheer2017deep}
M.~Zaheer, S.~Kottur \emph{et~al.}, ``Deep sets,'' \emph{Neural Inf. Proces.
  Syst. (NeurIPS)}, Long Beach, CA, United states, Dec. 2017.

\bibitem{ravanbakhsh2017equivariance}
S.~Ravanbakhsh, J.~Schneider, and B.~Poczos, ``Equivariance through
  parameter-sharing,'' in \emph{Int. Conf. Mach. Learn. (ICML)}, vol.~6,
  Sydney, NSW, Australia, Aug. 2017, pp. 2892--2901.

\bibitem{shen2022graph}
Y.~Shen, J.~Zhang \emph{et~al.}, ``Graph neural networks for wireless
  communications: {F}rom theory to practice,'' \emph{IEEE Trans. Wireless
  Commun.}, vol.~22, no.~5, pp. 3554--3569, May 2023.

\bibitem{yi2015topological}
X.~Yi and D.~Gesbert, ``Topological interference management with transmitter
  cooperation,'' \emph{IEEE Trans. Inf. Theory}, vol.~61, no.~11, pp.
  6107--6130, Nov. 2015.

\bibitem{shen2020graph}
Y.~Shen, Y.~Shi \emph{et~al.}, ``Graph neural networks for scalable radio
  resource management: {A}rchitecture design and theoretical analysis,''
  \emph{IEEE J. Sel. Areas Commun.}, vol.~39, no.~1, pp. 101--115, Jan. 2020.

\bibitem{kim2022bipartite}
J.~Kim, H.~Lee \emph{et~al.}, ``A bipartite graph neural network approach for
  scalable beamforming optimization,'' \emph{IEEE Trans. Wireless Commun.},
  vol.~22, no.~1, pp. 333--347, Jan. 2023.

\bibitem{guo2021learning}
J.~Guo and C.~Yang, ``Learning power allocation for multi-cell-multi-user
  systems with heterogeneous graph neural networks,'' \emph{IEEE Trans.
  Wireless Commun.}, vol.~21, no.~2, pp. 884--897, Feb. 2021.

\bibitem{zhao2022learning}
B.~Zhao, J.~Guo, and C.~Yang, ``Learning precoding policy: {CNN} or {GNN}?'' in
  \emph{IEEE Wireless Commun. Networking Conf. (WCNC)}, Austin, TX, USA, Apr.
  2022.

\bibitem{liu2023multidimensional}
S.~Liu, J.~Guo, and C.~Yang, ``Multidimensional graph neural networks for
  wireless communications,'' \emph{IEEE Trans. Wireless Commun.}, vol.~23,
  no.~4, pp. 3057--3073, Apr. 2024.

\bibitem{he2022joint}
S.~He, J.~Yuan \emph{et~al.}, ``Joint user scheduling and beamforming design
  for multiuser {MISO} downlink systems,'' \emph{IEEE Trans. Wireless Commun.},
  vol.~22, no.~5, pp. 2975--2988, May 2023.

\bibitem{guo2021survey}
H.~Guo, J.~Li \emph{et~al.}, ``A survey on space-air-ground-sea integrated
  network security in {6G},'' \emph{IEEE Commun. Surv. Tutor.}, vol.~24, no.~1,
  pp. 53--87, Firstquarter 2022.

\bibitem{9298921}
K.~Pratik, B.~D. Rao, and M.~Welling, ``{RE}-{MIMO}: Recurrent and permutation
  equivariant neural {MIMO} detection,'' \emph{IEEE Trans. Signal Process.},
  vol.~69, pp. 459--473, Dec. 2020.

\bibitem{wang2023soft}
Y.~Wang, H.~Hou \emph{et~al.}, ``Soft demodulator for symbol-level precoding in
  coded multiuser {MISO} systems,'' \emph{IEEE Trans. Wireless Commun.},
  vol.~23, no.~10, pp. 14\,819--14\,835, Oct. 2024.

\bibitem{hartford2018deep}
J.~Hartford, D.~Graham \emph{et~al.}, ``Deep models of interactions across
  sets,'' in \emph{Int. Conf. Mach. Learn. (ICML)}, vol.~5, Stockholm, Sweden,
  Jul. 2018, pp. 3050--3061.

\bibitem{keriven2019universal}
N.~Keriven and G.~Peyr{\'e}, ``Universal invariant and equivariant graph neural
  networks,'' \emph{Neural Inf. Proces. Syst. (NeurIPS)}, vol.~32, Vancouver,
  BC, Canada, Dec. 2019.

\bibitem{lee2019set}
J.~Lee, Y.~Lee \emph{et~al.}, ``Set transformer: {A} framework for
  attention-based permutation-invariant neural networks,'' in \emph{Int. Conf.
  Mach. Learn. (ICML)}, vol.~97, Long Beach, CA, United states, Jun. 2019, pp.
  3744--3753.

\bibitem{maron2018invariant}
H.~Maron, H.~Ben-Hamu \emph{et~al.}, ``Invariant and equivariant graph
  networks,'' in \emph{Int. Conf. Learn. Represent. (ICLR)}, New Orleans, LA,
  United states, May. 2019.

\bibitem{artin2011algebra}
\BIBentryALTinterwordspacing
M.~Artin, \emph{Algebra}.\hskip 1em plus 0.5em minus 0.4em\relax Pearson
  Education, 2011. [Online]. Available:
  \url{https://books.google.com.hk/books?id=S6GSAgAAQBAJ}
\BIBentrySTDinterwordspacing

\bibitem{ravanbakhsh2020universal}
S.~Ravanbakhsh, ``Universal equivariant multilayer perceptrons,'' in \emph{Int.
  Conf. Mach. Learn. (ICML)}, vol. PartF168147-11, Virtual, Online, Jul. 2020,
  pp. 7952--7962.

\bibitem{kim2022pure}
J.~Kim, D.~Nguyen \emph{et~al.}, ``Pure transformers are powerful graph
  learners,'' \emph{Neural Inf. Proces. Syst. (NeurIPS)}, vol.~35, pp.
  14\,582--14\,595, New Orleans, LA, United states, Nov. 2022.

\bibitem{maron2020learning}
H.~Maron, O.~Litany \emph{et~al.}, ``On learning sets of symmetric elements,''
  in \emph{Int. Conf. Mach. Learn. (ICML)}, vol. PartF168147-9, Virtual,
  Online, Jul. 2020, pp. 6690--6700.

\bibitem{zhang2022deep}
M.~Zhang, J.~Gao, and C.~Zhong, ``A deep learning-based framework for low
  complexity multiuser mimo precoding design,'' \emph{IEEE Trans. Wireless
  Commun.}, vol.~21, no.~12, pp. 11\,193--11\,206, 2022.

\bibitem{luo2016understanding}
W.~Luo, Y.~Li \emph{et~al.}, ``Understanding the effective receptive field in
  deep convolutional neural networks,'' \emph{Advances in neural information
  processing systems}, vol.~29, 2016.

\bibitem{pan2022permutation}
H.~Pan and R.~Kondor, ``Permutation equivariant layers for higher order
  interactions,'' in \emph{International Conference on Artificial Intelligence
  and Statistics (AISTATS)}, Valencia, Spain, Apr. 2023, pp. 5987--6001.

\bibitem{vaswani2017attention}
A.~Vaswani, N.~Shazeer, and N.~Parmar, ``Attention is all you need,'' in
  \emph{Neural Inf. Proces. Syst. (NeurIPS)}, Long Beach, CA, United states,
  Dec. 2017.

\bibitem{10198239}
Y.~Shi, L.~Lian \emph{et~al.}, ``Machine learning for large-scale optimization
  in {6G} wireless networks,'' \emph{IEEE Commun. Surv. Tutor.}, vol.~25,
  no.~4, pp. 2088--2132, Fourthquarter 2023.

\bibitem{mukhtar2012adaptive}
H.~Mukhtar and M.~El-Tarhuni, ``An adaptive hierarchical {QAM} scheme for
  enhanced bandwidth and power utilization,'' \emph{IEEE Trans. Commun.},
  vol.~60, no.~8, pp. 2275--2284, Aug. 2012.

\bibitem{8481590}
C.~Qian, X.~Fu \emph{et~al.}, ``Tensor-based channel estimation for
  dual-polarized massive {MIMO} systems,'' \emph{IEEE Trans. Signal Process.},
  vol.~66, no.~24, pp. 6390--6403, Dec. 2018.

\bibitem{he2016identity}
K.~He, X.~Zhang \emph{et~al.}, ``Identity mappings in deep residual networks,''
  in \emph{Eur. Conf. Comput. Vis. (ECCV)}, vol. 9908 LNCS, Amsterdam, The
  Netherlands, Oct. 2016, pp. 630--645.

\bibitem{ba2016layer}
\BIBentryALTinterwordspacing
J.~L. Ba, J.~R. Kiros, and G.~E. Hinton, ``Layer normalization,'' \emph{arXiv
  preprint arXiv:1607.06450}, 2016. [Online]. Available:
  \url{https://arxiv.org/abs/1607.06450}
\BIBentrySTDinterwordspacing

\bibitem{6758357}
S.~Jaeckel, L.~Raschkowski \emph{et~al.}, ``Quadriga: A {3-D} multi-cell
  channel model with time evolution for enabling virtual field trials,''
  \emph{IEEE Trans. Antennas Propag.}, vol.~62, no.~6, pp. 3242--3256, Mar.
  2014.

\bibitem{3GPPTR38.901}
\textit{Technical Specification Group Radio Access Network; Study on channel
  model for frequencies from 0.5 to 100 GHz (Release 16)}, document 3GPP TR
  38.901, Version 16.1.0, 3rd Generation Partnership Project Dec. 2019.

\bibitem{5673745}
V.~Raghavan and A.~M. Sayeed, ``Sublinear capacity scaling laws for sparse
  {MIMO} channels,'' \emph{IEEE Trans. Inf. Theory}, vol.~57, no.~1, pp.
  345--364, Jan. 2011.

\bibitem{kingma2014adam}
\BIBentryALTinterwordspacing
D.~P. Kingma and J.~Ba, ``Adam: A method for stochastic optimization,''
  \emph{arXiv preprint arXiv:1412.6980}, 2014. [Online]. Available:
  \url{https://arxiv.org/abs/1412.6980}
\BIBentrySTDinterwordspacing

\bibitem{1391204}
C.~Peel, B.~Hochwald, and A.~Swindlehurst, ``A vector-perturbation technique
  for near-capacity multiantenna multiuser communication-{P}art {I}: {C}hannel
  inversion and regularization,'' \emph{IEEE Trans. Commun.}, vol.~53, no.~1,
  pp. 195--202, Jan. 2005.

\end{thebibliography}

\end{document}